\documentclass[fleqn,usenatbib]{mnras}

\usepackage{newtxtext,newtxmath}

\usepackage[T1]{fontenc}

\DeclareRobustCommand{\VAN}[3]{#2}
\let\VANthebibliography\thebibliography
\def\thebibliography{\DeclareRobustCommand{\VAN}[3]{##3}\VANthebibliography}

\usepackage{graphicx}	
\usepackage{amsmath}	
\usepackage{subcaption}
\usepackage{siunitx} 
\usepackage{hyperref}
\usepackage{xcolor} 
\usepackage{textcomp}

\newcommand {\Msun} {M$_\odot$}

\defcitealias{valenti_near-infrared_2010}{Val10}

\title{Hunting for intermediate-mass black holes in globular clusters: an~astrometric study of NGC~6441}

\author[Häberle et al.]{
Maximilian Häberle$^{1}$\thanks{E-mail: haeberle@mpia.de},
Mattia Libralato$^{2}$,
Andrea Bellini$^{3}$,
Laura L. Watkins$^{2}$,
Jörg-Uwe Pott$^{1}$,
\newauthor
Nadine Neumayer$^{1}$,
Roeland P. van der Marel$^{3,\,4}$,
Giampaolo Piotto$^{5,\,6}$, and
Domenico Nardiello$^{7,\,6}$
\\
$^{1}$Max-Planck-Institut für Astronomie, Königstuhl 17, 69117 Heidelberg, Germany\\
$^{2}$AURA for the European Space Agency (ESA), ESA Office, Space Telescope Science Institute, 3700 San Martin Drive, Baltimore MD 21218, USA\\
$^{3}$Space Telescope Science Institute, 3700 San Martin Drive, Baltimore MD 21218, USA\\
$^{4}$Center for Astrophysical Sciences, Department of Physics \& Astronomy, Johns Hopkins University, Baltimore, MD 21218, USA\\
$^{5}$Dipartimento di Fisica e Astronomia ``Galileo Galilei'', Universit\`{a} degli Studi di Padova, Vicolo dell'Osservatorio 3, I-35122 Padova,
Italy\\
$^{6}$Istituto Nazionale di Astrofisica (INAF), Osservatorio Astronomico di Padova, Vicolo dell'Osservatorio 5, I-35122 Padova, Italy\\
$^{7}$Aix Marseille Univ, CNRS, CNES, LAM, Marseille, France
}

\date{Accepted XXX. Received YYY; in original form ZZZ}

\pubyear{2020}

\begin{document}
\label{firstpage}
\pagerange{\pageref{firstpage}--\pageref{lastpage}}
\maketitle

\begin{abstract}
We present an astrometric study of the proper motions (PMs) in the core of the globular cluster NGC~6441. The core of this cluster has a high density and observations with current instrumentation are very challenging. We combine ground-based, high-angular-resolution NACO@VLT images with \textit{Hubble Space Telescope} ACS/HRC data and measure PMs with a temporal baseline of 15~yr for about 1400 stars in the centermost 15 arcseconds of the cluster. We reach a PM precision of~$\sim$30~\textmu as~yr$^{-1}$ for bright, well-measured stars.

Our results for the velocity dispersion are in good agreement with other studies and extend already-existing analyses of the stellar kinematics of NGC~6441 to its centermost region never probed before. In the innermost arcsecond of the cluster, we measure a velocity dispersion of (19.1~$\pm$~2.0)~km~s$^{-1}$ for evolved stars. Because of its high mass, NGC~6441 is a promising candidate for harbouring an intermediate-mass black hole (IMBH). We combine our measurements with additional data from the literature and compute dynamical models of the cluster. We find an upper limit of $M_{\rm IMBH} < 1.32 \times 10^4\,\textrm{M}_\odot$ but we can neither confirm nor rule out its presence. We also refine the dynamical distance of the cluster to $12.74^{+0.16}_{-0.15}$~kpc. 

Although the hunt for an IMBH in NGC~6441 is not yet concluded, our results show how future observations with extremely-large telescopes will benefit from the long temporal baseline offered by existing high-angular-resolution data.
\end{abstract}




\section{Introduction}
Globular clusters are the oldest surviving stellar systems in the Galaxy. Because of their long dynamical timescales, they are witnesses of the early history of the Milky Way. 
The study of the internal dynamics of globular clusters is an active field of research and can reveal information about the formation and evolution of the clusters themselves, their interaction with the Galactic potential, but also the possible existence of intermediate-mass black holes (IMBHs) in their core.

IMBHs have been predicted in the centres of globular clusters \citep{portegies_zwart_runaway_2002, miller_production_2002}, but a definitive detection is still lacking (see \citealt{greene_intermediate-mass_2019} for a recent review). As the sphere of influence of a hypothetical IMBH is limited to the very centre of the cluster, dynamical studies of this central region are necessary to infer the presence of an IMBH.

There are two observational methods to study the kinematic signature of individual stars in a globular cluster: spectroscopic line-of-sight (LOS) velocity measurements and astrometric measurements of proper motions (PMs). While in the early days of the field the line-of-sight velocities of only small numbers of stars were measured, in recent years significant progresses have been made both in spectroscopy and astrometry.
The MUSE integral field spectrograph allowed LOS velocity measurements of up to 20,000 stars within the half-light radius of a large sample of globular clusters \citep{kamann_stellar_2018}.

On the astrometry side, the best tool to study the crowded cores of the clusters is the \textit{Hubble Space Telescope (HST)} and there are catalogues with high precision PM measurements for up to a hundred thousand stars \citep{bellini_hubble_2014, libralato_hubble_2018} in a single cluster. Additionally, the \textit{Gaia} satellite \citep{2016A&A...595A...2G} allows the study of stellar motions in the outskirts of globular clusters. However, its completeness and precision are limited in cluster centres, where the stellar density is high.

Observations of the crowded cores of some globular clusters still remain very challenging with current instrumentation. Crowding effects limit the usability of spectroscopic facilities, therefore, astrometry with high resolution imagers is the method of choice. One such example is the globular cluster NGC~6441: its core is extremely crowded and neither of the current \textit{HST} imagers provides the necessary resolution for astrometric studies of the core. The Advanced Camera for Surveys  Wide Field Channel (ACS/WFC) has a pixel scale of 50~mas~pixel$^{-1}$ and the Wide Field Camera 3 (WFC3) UVIS has a pixel scale of 40~mas~pixel$^{-1}$.

The High Resolution Channel of the Advanced Camera for Surveys (ACS/HRC) had a smaller pixel scale of 25~mas~pixel$^{-1}$ and, while still affected by the high stellar density, could measure stellar position in the core of NGC~6441 with high precision. A first epoch of observations was taken in 2003, but the failure of this instrument in 2006 made it impossible to observe the core of NGC~6441 again. 

Since there are no other suitable epochs of \textit{HST} observations, we re-observed the core of the cluster using the near infrared instrument NACO at the Very Large Telescope (VLT) of the European Southern Observatory (ESO). The adaptive-optics (AO)-assisted observations close to the diffraction limit with an 8-m class telescope in the $K_{\rm S}$ band have a point spread function (PSF) full width half maximum (FWHM) of about 73~mas, similar to that of the \textit{HST} at visible wavelengths, and with 13.2~mas~pixel$^{-1}$ the pixel scale is twice as small as the one of ACS/HRC.

NGC~6441 is a Galactic bulge cluster. With a mass of \num{1.2e6}~$\text{M}_\odot$ \citep{baumgardt_catalogue_2018}, it is one of the most massive clusters in the Galaxy. Furthermore, it is metal rich ([Fe/H] $= -0.55$, \citealt{harris_new_2010}) and hosts at least two main stellar populations \citep{bellini_hubble_2014}.

The dynamics of the cluster have been studied in several papers, both with PMs \citep{watkins_hubble_2015} and LOS velocities \citep{kamann_stellar_2018}, but due to its high density, accurate measurements of the velocity dispersion of the stars in the very centre of the cluster are still lacking. Our study probes the velocity dispersion to unprecedentedly small radii. While we cannot put strong constraints on the presence of an IMBH, we add more than 20 PM measurements in the innermost arcsecond. The combination of space-based and ground-based AO-assisted astrometric imaging with a long time baseline is a successful pilot study to showcase what will be possible with new instrumentation expected in the next decade, such as the MICADO imager at the Extremely Large Telescope \citep{2016SPIE.9908E..1ZD}.

This paper is divided in the following sections: first we describe our observations (Section \ref{sec:obs}), after which we present the data analysis in Section \ref{sec:data}. The determination of the PMs is described in Section \ref{sec:pm}. We use our results to create dynamical models of the globular cluster in Section \ref{sec:kin}, and discuss and conclude our work in Section \ref{sec:con}.


\section{Observations}
\label{sec:obs}
\subsection{Epoch 1: 2003 \textit{HST} ACS/HRC Observations}
The \textit{HST} instrument ACS/HRC was in operation from 2002 to 2006. It featured a $1024\times1024$~pixel CCD detector with a pixel scale of $28\times25$~mas$^2$pixel$^{-1}$, giving it a field of view of $\sim 29\times26$~arcsec$^2$.

The core of NGC~6441 was observed with the ACS/HRC in two broadband filters (F555W and F814W) during Program GO-9835 (PI: G. Drukier). The 36 exposures in the F555W filter have all the same exposure time of 240 s. The F814W band exposures comprise of 5 short and 12 long exposures of 40~s and 440~s, respectively.
\subsection{Epoch 2: 2018 NACO @ VLT Observations}
\subsubsection{The instrument}
NACO (short for NAOS-CONICA) was a near-infrared, adaptive-optics-assisted imager and spectrograph at the Very Large Telescope (VLT). A full description of the instrument and its performance can be found in \cite{lenzen_naos-conica_2003} and \cite{rousset_naos--first_2003}.  The instrument was mounted on the Nasmyth B focus of UT4 from 2001 to 2013, and then moved to UT1 in 2014, where it continued its operations until its recent decommissioning in October 2019.
The adaptive optics front end (NAOS) was equipped with a 185~actuator deformable mirror, a tip/tilt mirror and two wavefront sensors (operating in the visual and the IR range). Using bright natural guide stars, the best Strehl ratio obtainable in the $K_{\rm S}$ band was around 50\% in typical observing conditions, similar to what we achieved in our observing run.

The camera CONICA was equipped with an Aladdin2 detector (1026 x 1024 pixels, InSb) that replaced the Aladdin3 detector that was in use from 2004 to 2013. The detector was affected by several artefacts that are described in detail later. CONICA offered cameras with three different pixel scales (S13: 13.22~mas~pixel$^{-1}$, S27: 27.06~mas~pixel$^{-1}$, S54: 54.3~mas~pixel$^{-1}$). For our observations of the core of NGC~6441, we made use of the S13 mode to obtain the highest possible resolution.
\subsubsection{The dataset}
\begin{figure*}
  \centering
    \includegraphics[width=1.0\textwidth]{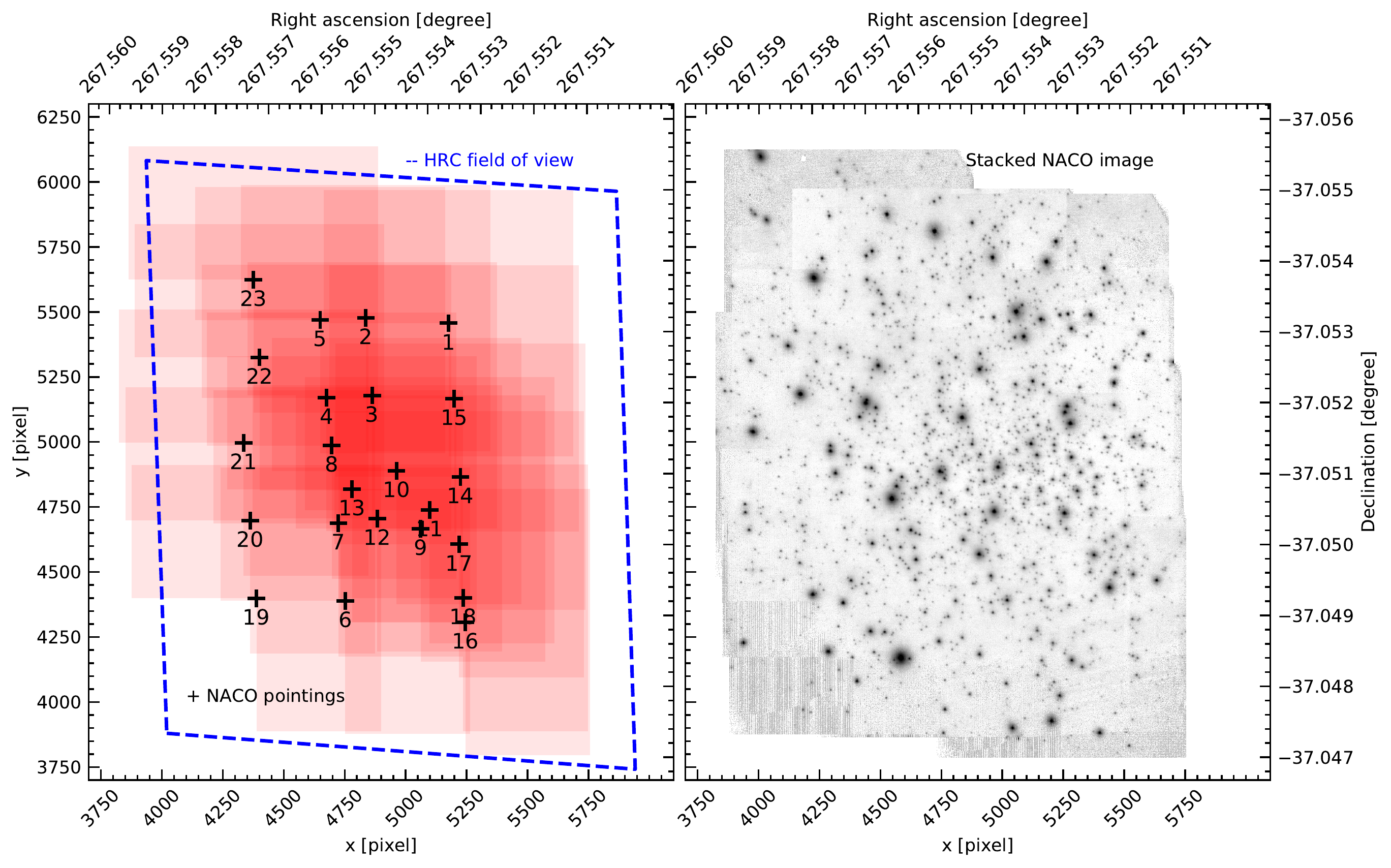}
  \caption{The two panels show the field of the NACO observations in equatorial coordinates and in our pixel-based coordinate system. In the left plot, the centres of all long NACO pointings are marked with a black cross and the footprint of each exposure is marked in red. The missing lower-left quadrant is already removed from the footprints. In the central regions, a maximum of 17 exposures overlap, while at the edges of the observed field the depth of coverage is much smaller. The right plot shows a stacked image of the NACO observations in an inverted grey scale.}
  \label{fig:finding_chart}
\end{figure*}

The second epoch of observations of the core of NGC~6441 were obtained with the ESO Program ID 0101.D-0385 (PI: M. Libralato).
We only made use of images taken during the August run because of their overall better quality. In these nights, there were 23 usable exposures of \SI{200}{\second} each, and 27 short observations of \SI{30}{\second} each. The observations were executed with a dither pattern that covered the field of the first-epoch \textit{HST} observation almost completely (see \autoref{fig:finding_chart}). The observing strategy was designed to solve for the geometric distortion of the detector using an auto-calibration approach \citep[see, e.g.,][]{libralato_ground-based_2014} and to achieve an astrometric precision high enough to allow for the kinematic analysis of NGC~6441.

\begin{table*}
\centering
\begin{tabular}{lllllll}
\hline
Epoch & Telescope & Instrument & Program ID & Filter & $t_{\text{exp}}$ & $N_{\text{images}}$ \\\hline
\hline
2003.67      & \textit{HST}        & ACS/HRC    &   GO-9835         & F555W  & 200 s     & 36   \\
             &                     &            &                   & F814W  & 440 s     & 12   \\
             &                     &            &                   & F814W  & 40 s      & 5    \\
\hline
2018.61      & VLT                 & NACO       &   0101.D-0385     & $K_{\rm S}$     & 200 s     & 23   \\
             &                     &            &                   &        & 30 s      & 27        \\\hline
\end{tabular}
\caption{List of used observations.}
\end{table*}

\section{Data analysis}
\label{sec:data}
PM measurements require multiple epochs of precise stellar position measurements. In this section we describe how we extracted stellar positions from the raw exposures of our two datasets.
\subsection{Analysis of Epoch-1 data (\textit{HST} ACS/HRC 2003)}
\label{da_hst}
The data reduction of the ACS/HRC exposures was performed on \texttt{\_flt}-type images\footnote{These images are bias, dark and flat-field corrected by the standard \textit{HST} pipeline \texttt{CALACS}, but are not resampled, so they retain the full signal of the astrometric scene.} by closely following the prescriptions given in \cite{bellini_hubble_2014, bellini_state---art_2017}.

In brief, we started by deriving state-of-art, spatially-variable PSF models for each exposure by perturbing the library PSF models created by Jay Anderson\footnote{\url{https://www.stsci.edu/~jayander/STDPSFs/ACSHRC/}}. Our improved PSF models take into account telescope breathing effects (\citealt{2008acs..rept....3D}, see also \citealt{bellini_state---art_2017, 2018acs..rept....8B}), which can significantly change the shape of the \textit{HST} PSFs from one exposure to the next even within the same telescope orbit.

Preliminary stellar positions and fluxes of bright sources were obtained through PSF fitting using the \texttt{FORTRAN} code \texttt{hst1pass} (Anderson in preparation, see \citealt{2018ApJ...853...86B} for details). Photometry of saturated stars includes all the relevant flux that has bled into adjacent pixels following the prescriptions given in \cite{2010wfc..rept...10G}. Stellar positions were corrected for the effects of geometric distortion using the state-of-the-art solutions provided by \cite{2004acs..rept...15A}.

Next, we made use of \textit{Gaia} DR2 positions (\citealt{2016A&A...595A...2G, 2018A&A...616A...1G}) to define a reference-frame system with North up, East to the left, and with the same pixel scale of NACO of 13.2 mas\,pixel$^{-1}$. We transformed stellar positions of each single-exposure catalogue on to the reference frame by means of general, six-parameter linear transformations. Our best estimate of positions and fluxes for all possible sources in the ACS/HRC field is obtained using the \texttt{FORTRAN} code \texttt{KS2} (Anderson in preparation, see \citealt{bellini_state---art_2017} for details). \texttt{KS2} starts from the image-tailored PSF models, the lists of bright stars and their transformations on to the reference frame, and goes through several waves of source finding, measuring and subtraction using all the exposures simultaneously. Our final first-epoch catalogue contains around 44\,000 sources measured in both F555W and F814W filters. In addition to the photometry results, the output contains several quality parameters such as the radial excess value (\texttt{RADXS}). If it is positive, the profile of a single star contains excessive flux outside of the fit radius with respect to the PSF, if it is lower than 0, this flux is lower than expected.

To calibrate the photometry of the ACS/HRC data, we performed aperture photometry on the corresponding \texttt{\_drz} images that are resampled and normalised to 1 s exposure time, corrected the results for finite aperture (using the corrections from \citealt{bohlin_perfecting_2016}) and then brought them onto the VegaMAG system using the zeropoints available at the STScI website\footnote{\url{https://acszeropoints.stsci.edu/}}. Then we determined the zeropoint between the calibrated aperture photometry and our PSF photometry by taking the 3$\sigma$-clipped median of the magnitude difference for bright isolated stars (we chose stars that had no brighter neighbours within a 18 pixel radius). A similar calibration process for \textit{HST} photometry has been described e.g. in \cite{bellini_state---art_2017}.
\subsection{Analysis of Epoch-2 data (NACO@VLT 2018)}
\subsubsection{Pre-reduction}
We downloaded the raw NACO exposures and the corresponding calibration frames (dark frames, flat-fields, bad-pixel maps) from the ESO Science archive. The dark frames were not usable because the background showed patterns that varied over the course of the night. We therefore only divided the images by the flat-fields.
Furthermore, we flagged all saturated pixels and added them to the bad-pixel map.  The saturation threshold is set at 10\,000 analogue-digital converter units (ADUs) to avoid non-linearity effects following the recommendations of the NACO User manual \citep{schmidtobreick_naco_2018}. The bottom-left quadrant of NACO presents a high number of bad columns in a regular pattern (2 good, 2 bad, 3 good, 1 bad), so we chose not to use it in our analysis.
\subsubsection{Background Model}
\begin{figure}
\centering
\includegraphics[width=0.5\textwidth]{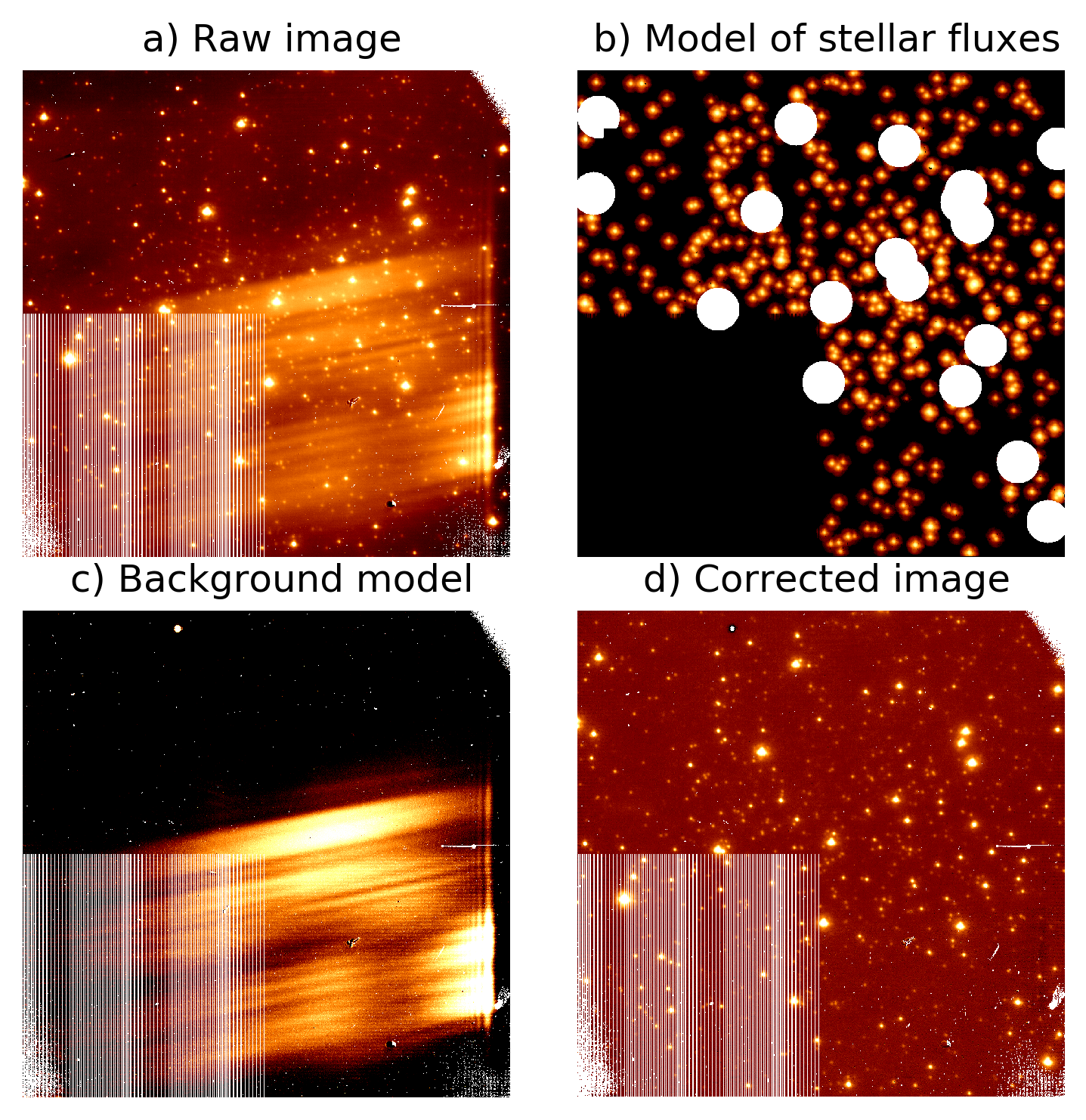}
\caption{The top-left panel (a) shows a typical raw NACO image of NGC~6441. The thermal-related background pattern and the bad columns in the bottom-left quadrant are clearly visible. In the top-right panel (b), we present the corresponding image model. Each star in the model image is obtained by rescaling the PSF by the stellar flux. The white circles have a radius of 45~NACO~pixels and mark the region around saturated stars that we flagged and excluded from the model creation. The bottom-left image (panel c) is the resulting background model obtained after our iterative procedure. Finally, in the bottom-right panel (d), we show the image from panel (a) after the removal of the background pattern.}
\label{fig:background_model}
\end{figure}

Insufficient thermal shielding within the instrument\footnote{This effect is known since 2015, see ESO instrument history:\\\url{https://www.eso.org/sci/facilities/paranal/decommissioned/naco/History.html}} created a diffuse, non-static background pattern that cannot be corrected in the pre-reduction phase. The time-dependent nature of the background pattern and the high density of stars in the cluster centre made it challenging to correct this pattern. The correction we applied is the result of two iterations. In each iteration, we used the median of multiple images in which the background pattern did not change as our background model. To remove the influence of stars, we flagged pixels from an image if they were significantly brighter than in the other images. Furthermore, we flagged a circular area with $r=45$~NACO~pixels around saturated stars to remove the influence of their bright and extended halos. Flagged pixels were excluded from the median calculation. In the first iteration this clipping was performed directly on the raw images. In the second iteration we improved the model by subtracting the modelled stellar fluxes from the raw images (based on the PSF photometry results from the next paragraph) before running a jackknife algorithm to find additional outliers. For some images it was beneficial to shift the model in the $y$-direction before subtracting it. The different steps are visualised in \autoref{fig:background_model}.
\subsubsection{Determination of the PSF}
An accurate PSF model is a crucial ingredient for high-precision astrometry. We created a $3\times3$ array of empirical PSFs to take into account the spatial variation of the PSF caused by the telescope optics and AO-related setup. Since the AO performance and seeing conditions change over time, we tailored these PSF arrays to each image.
This method was introduced for ground-based instruments in \cite{anderson_ground-based_2006} and has also been adapted to various near-infrared instruments \citep{libralato_ground-based_2014, libralato_high-precision_2015, kerber_deep_2019}.  We refer to these papers for a detailed description of the PSF modelling. Here we provide only a brief overview of the method and discuss the major differences with respect to the original papers.

\begin{figure}
  \centering
    \includegraphics[width=0.25\textwidth]{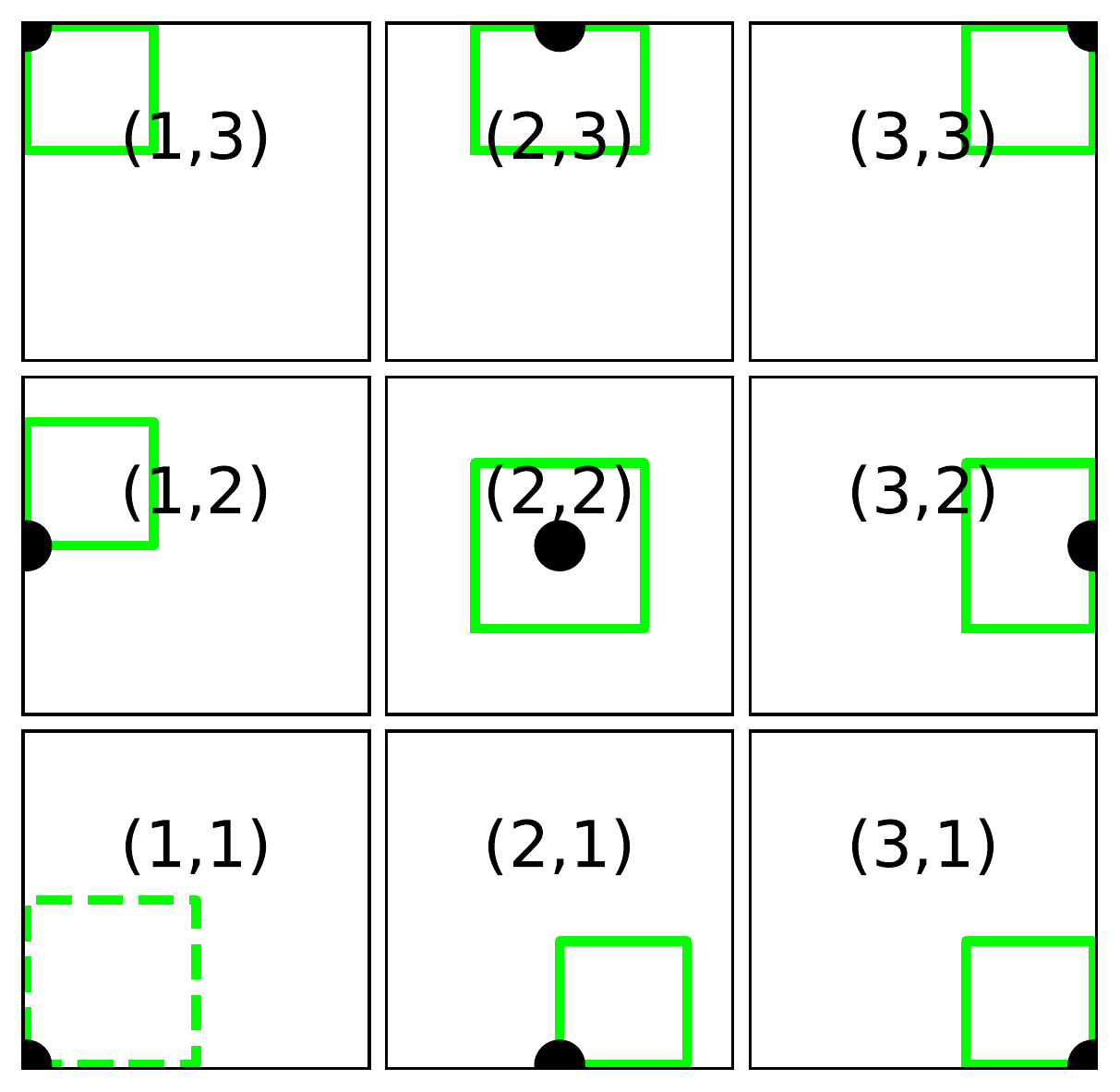}
    \includegraphics[width=0.49\textwidth]{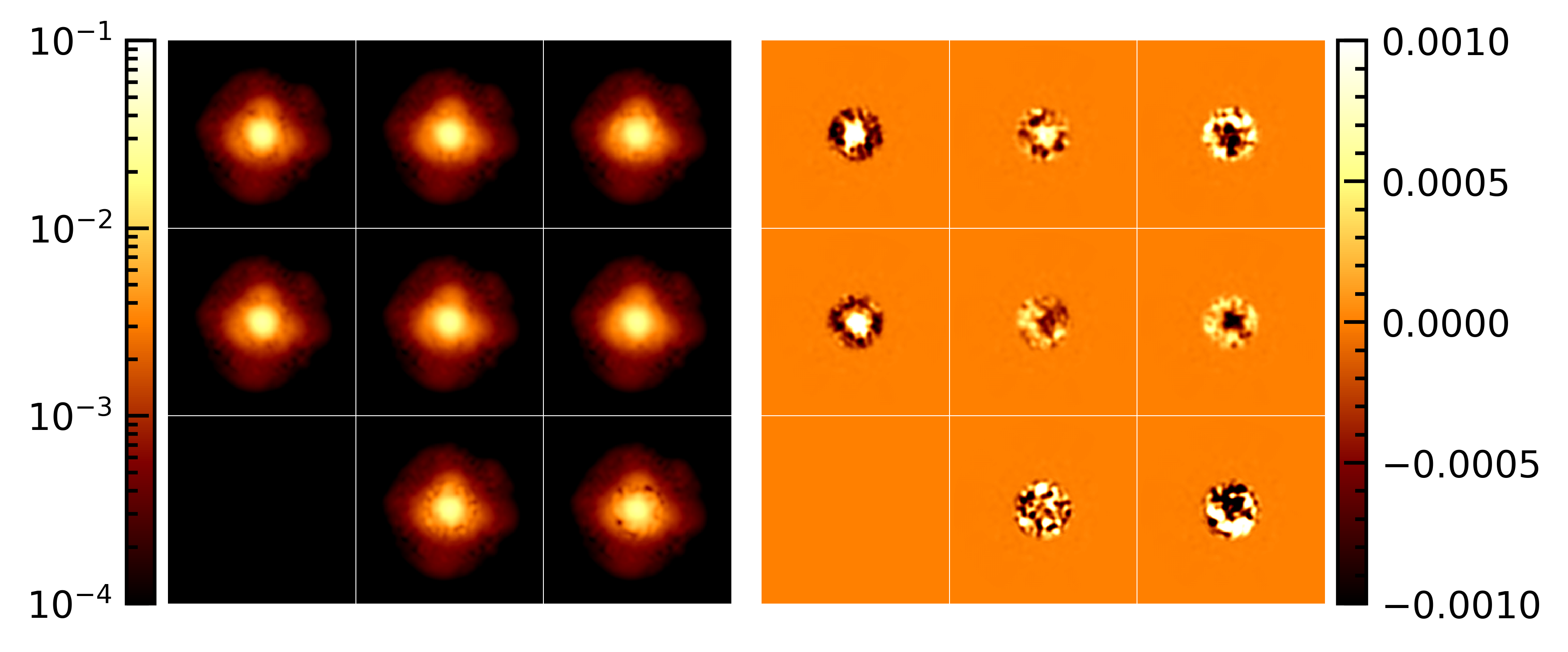}
      \caption{(Upper panel:) Each of the nine squares is a representation of the detector. Each black point shows the location of a local PSF model that is determined from stars within the green rectangle. (Lower-left panel:) The $3\times3$ grid of PSFs determined for one NACO frame. (Lower-right panel:) Residuals between each PSF and one average PSF for the entire frame. Darker/lighter regions represent an excess/deficit of flux of the spatially-variable PSF with respect to the average PSF. }
  \label{fig:psf}
\end{figure}

Because of the small field of view of NACO, we used a regular $3\times3$ grid of PSF models to map the spatial variability. Bilinear interpolation between the grid points was used to determine the PSF on each location of the detector. Due to the unusable lower-left quadrant, the lower-left grid point contains no PSF. For every other gridpoint, the local PSF was determined from at least 20 well-measured, isolated bright stars in a nearby rectangular cell. The exact layout of these cells and the gridpoint locations are shown in \autoref{fig:psf}.

To determine the Strehl ratio of our observations, we divided the central value of our different PSF models by the central value of an Airy function with a radius of 69 mas (or 5.2~NACO~pixels), the expected diffraction-limited PSF for a 8-m telescope observing at 2.2 \textmu m wavelength. The Strehl ratios we obtained ranged between 0.2 and 0.49, with a median value of 0.37. This is compatible with the typical Strehl ratios stated in the NACO User manual. The median FWHM of our PSF was 73~mas. \autoref{fig:psf} shows the 8 PSF models determined for an individual frame and how the local models differ from a spatially constant PSF model.
\subsubsection{Fitting the stellar positions}
After the PSF model has been determined for each image, we used it to fit the raw position (x,y) in pixels and the instrumental magnitude of each star. To be able to select well~measured stars, we also calculated the so called quality-of-fit (\texttt{QFIT}) value, which is a normalised sum of the fit residuals within the fitting radius \citep{anderson_ground-based_2006}. The closer to 0 the \texttt{QFIT}, the better is the PSF fit.

\subsubsection{Geometric distortion correction (GDC)}
To achieve a sub-pixel astrometric precision, we have to correct the geometric distortion present in the NACO images, which reaches up to 2~NACO~pixels in the corners of the detector. We redetermined the GDC using our \textit{HST} catalogue as distortion-free reference, the process is described in detail in the Appendix \ref{app}. We corrected the geometric distortion to a level of $\le 0.03$~NACO~pixel ($\approx 0.4 \text{ mas}$).

In addition to the GDC using an external reference, we also employed a so called autocalibration approach in which the distortion free reference is obtained from the NACO data itself. However, this lead to a worse correction. The failure of the autocalibration approach with our 2018 NACO dataset highlights the importance of the choice of the dither $+$ rotation pattern, if such a calibration is required.
\subsubsection{Creation of a NACO master frame and tests on the astrometric precision}
Most stars in our dataset have been measured on multiple NACO exposures. To combine the distortion-corrected single-image catalogues, we matched them on the same reference system as defined for the first epoch (see \ref{da_hst}). We transformed the stellar positions from each distortion-corrected single-image catalogue on to the reference system using six-parameter linear transformations. Stars that have been measured in at least three images were then added to our NACO master frame. Our best estimate for their position is the averaged position from the individual transformed positions.

\begin{figure*}
  \centering
    \includegraphics[width=1.0\textwidth]{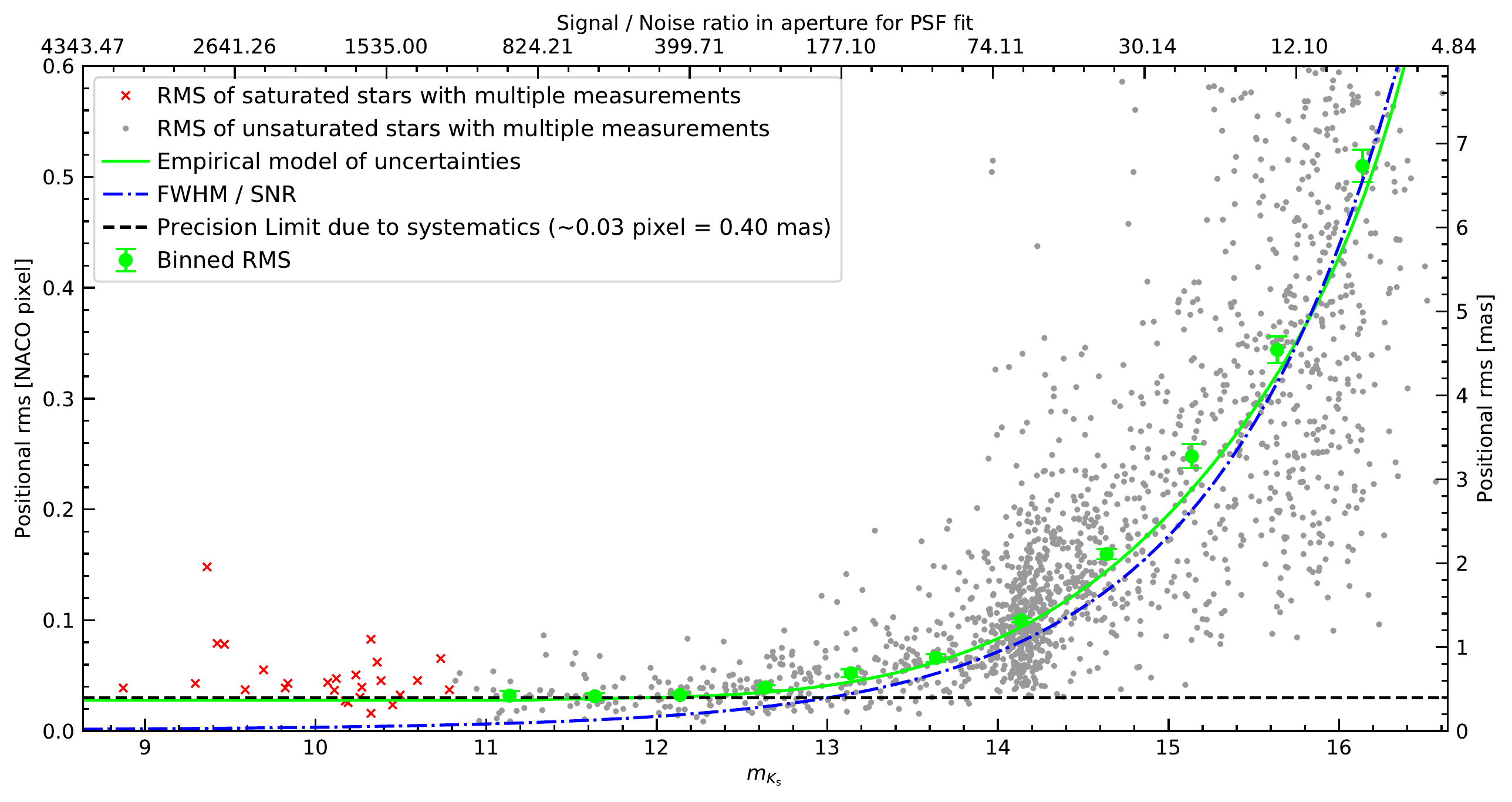}
  \caption{This plot shows the magnitude dependence of the astrometric precision reached with single, long (200 s) NACO exposures. We determine the positional scatter of stars that have been measured in at least 3 images by determining the rms of single-image position measurements when transformed in our reference system. Overall, faint stars follow the theoretical prediction for centroiding accuracy, while the measurements of bright stars are limited by a noise floor of $\sim$0.03~NACO~pixel.}
  \label{fig:precision_plot}
\end{figure*}

The scatter of the single measurements, quantified by their root mean square (rms) deviation from the master frame position, is a measure of the astrometric precision we can reach for a given signal to noise ratio (S/N). In \autoref{fig:precision_plot}, this positional rms is plotted as a function of the $m_{K_{\rm S}}$ magnitude (see subsection \ref{zeropoint} for our flux calibration). As expected, we see a decrease in precision for faint stars, i.e., those with a lower S/N ratio. For stars brighter than $m_{K_{\rm S}} < 12.5$, the trend flattens out at a level of approximately 0.03~NACO pixel ($\approx 0.40 $~mas), while theoretical limits on centroiding precision \citep[e.g.][]{lindegren_photoelectric_1978} predict:
\begin{equation}
    \sigma_\text{1D}=k\cdot\frac{\text{FWHM}_\text{PSF}}{\text{SNR}}
\end{equation}
(with $k$ close to unity).

This fundamental limit of 0.03~pixel is worse than the 0.01~pixel achieved for astrometry with \textit{HST} instruments, but comparable to other ground-based IR studies \citep{libralato_ground-based_2014, libralato_high-precision_2015, kerber_deep_2019}. It is most likely caused by systematic effects, such as residual uncorrected distortion, imperfections in our PSF models and/or systematic effects related to the thermal background pattern and its correction.
\subsubsection{Photometric calibration and creation of colour-magnitude diagrams}
\label{zeropoint}
\begin{figure}
  \centering
    \includegraphics[width=0.5\textwidth]{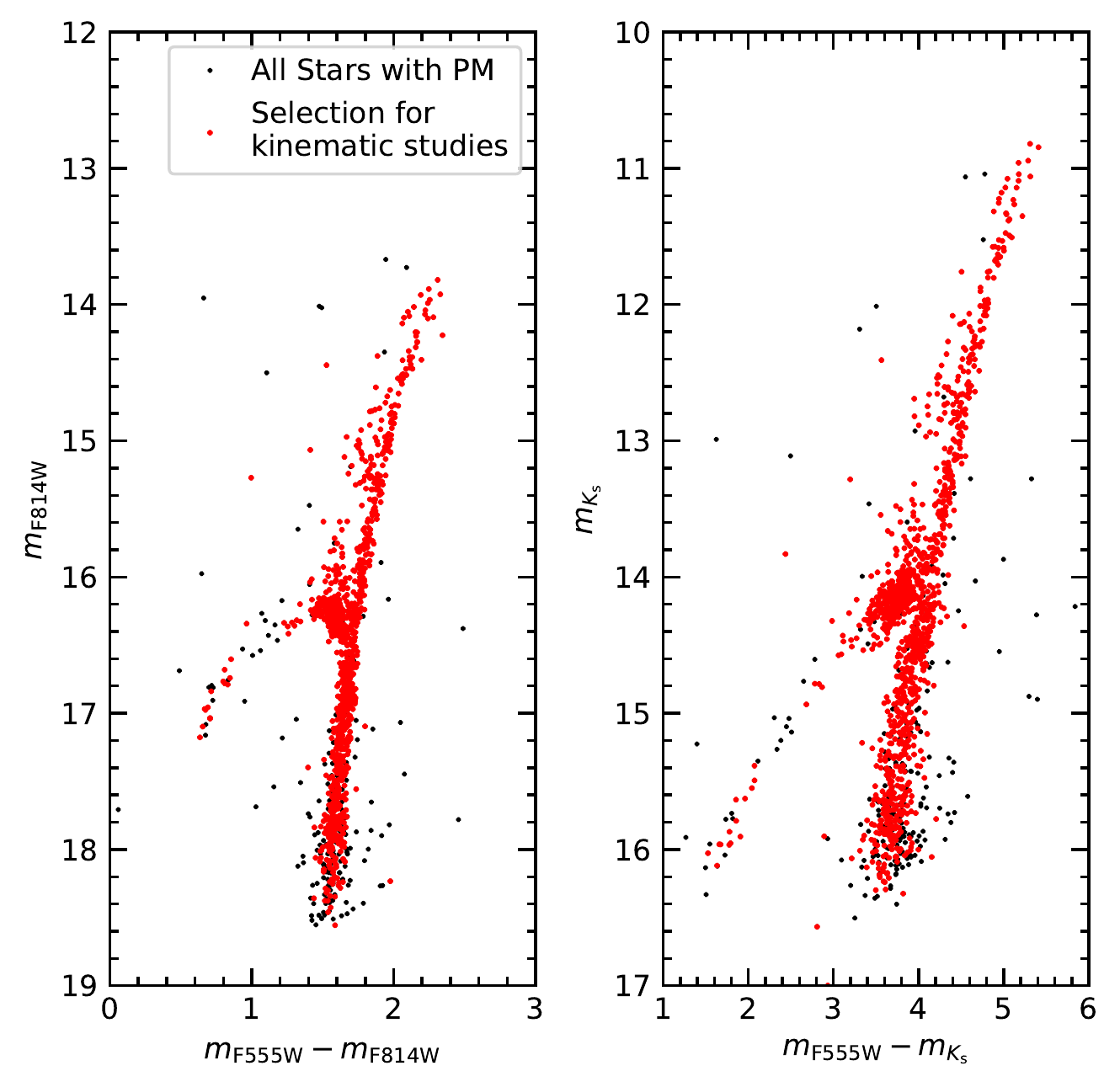}
      \caption{This plot shows two colour-magnitude diagrams of our PM sample based on the photometry of the ACS/HRC (F555W, F814W) and the NACO ($K_{\rm S}$) measurements. Measurements, that pass the quality selection for the kinematic analysis (Section \ref{sec:kin}) are marked in red.}
 \label{fig:hrc_cmd}
\end{figure}

We brought our NACO PSF photometry on to the Two Micron All-Sky Survey \citep[2MASS,][]{skrutskie_two_2006} photometric system by cross-matching it with the NGC~6441 catalogue published by \citet{valenti_near-infrared_2010} (in the following Val10). Due to the depth of the NACO images, most of the stars found in \citetalias{valenti_near-infrared_2010} are saturated in our long exposures. Therefore, we used a two step-process. First, we created a catalogue based on the short NACO exposures whose photometry has been zeropointed on to our long-exposure master-frame. Then, we determined the magnitude difference between the 63 stars in common between our master frame and the \citetalias{valenti_near-infrared_2010} catalogue. Our best estimate of the zeropoint is the sigma clipped median of the difference between the magnitudes in both catalogues.
Since we only have one filter, we added this simple zeropoint without accounting for possible colour effects.
After obtaining calibrated photometry for both the \textit{HST} and the NACO dataset, we created ($m_{\rm F555W}-m_{\rm F814W}$) and ($m_{\rm F555W}-m_{K_{\rm S}}$) colour-magnitudes diagrams, which can be seen in \autoref{fig:hrc_cmd}.

\section{Proper Motions}
\label{sec:pm}
To measure PMs one has to determine by how much the positions of individual stars have changed over time.

Compared to other studies, where multiple different epochs and fields of views are combined, the situation in our study is rather simple: we only combine two single epochs with a large time span in between. For this reason, we do not use a linear fit through multiple datapoints for each star, but we determine the position difference between two single epoch master frames. Due to the more reliably determined PSF and the better general astrometric precision, we decided to use only the long NACO exposures for this part of the analysis, at the cost of being unable to determine the PMs for around 20 stars, which were saturated in the long exposures.

\subsection{Creating the single-epoch master frames}
The goal of this step is to create new, improved master frames by using additional clipping procedures and local corrections. The following steps were performed on both the \textit{HST} ACS/HRC and the NACO data. 
\begin{description}
    \item[1. Selection of well-measured stars:] we applied several selections based on quality criteria (\texttt{QFIT} value, \texttt{RADXS} value) to restrict our sample to well-measured stars. Those well-measured stars were then used to determine the parameters of the following transformations.
    \item[2. Global transformations:] for each image, we use all stars flagged as ``well-measured'' to determine the optimal six-parameter linear transformations to transform the single-image catalogues on to the master frame. The residuals between the single-image catalogues and the master frame are used in the next steps for a local correction.
    \item[3. Local corrections:] after the parameters of the global transformations have been determined, we used local corrections to remove residual local effects such as uncorrected distortion. For each star, we determined the clipped mean of the transformation residuals of the 50 closest neighbours in both the $x$ and the $y$ directions. These mean residuals are subtracted from the mean coordinates of the star. This procedure is called ``boresight'' correction \citep{van_der_marel_new_2010}.
    \item[4. Error-based clipping and determination of mean positions:] the last steps gave us a list of multiple position measurements for each star, all transformed in the same reference frame. We defined as improved master-frame positions the mean positions of these locally-corrected measurements. Outliers were removed using a jackknife approach: for each star we excluded one measurement at a time and checked whether the excluded measurement deviates more than 10 standard deviations from the mean value of the remaining measurements. To determine the expected standard deviation, we employed the empirical error model based on the typical positional rms at a given magnitude (see \autoref{fig:precision_plot}). After the clipping process, we calculated the mean of the positions based on the remaining measurements. As the single-epoch position error, we employed the standard error of the mean for both coordinates: $\Delta x,y_{\rm master\,frame}=\frac{\sigma_{x,y}}{\sqrt{N}}$
\end{description}

\subsection{Combination of single-epoch master frames and determination of PMs}

The procedures used to match the two single-epoch master frames are similar to the methods used for the master frame creation. The difference is that, instead of matching single-image catalogues from the same epoch, we now directly match the master frame positions of the two different epochs. Since we want to measure  PMs relative to the bulk motion of the cluster, we use only bona-fide cluster members to determine the parameters of the linear transformations between the master frames.

\begin{description}
    \item[1. Selection of well-measured \textit{cluster members}:] we use the same quality selection as described above. In addition, we included a selection of bona-fide cluster members based on their location on the ACS/HRC based ($m_\textrm{F555W}-m_\textrm{F814W}$) CMD (see \autoref{fig:hrc_cmd}). Once PMs have been determined, we also restricted the sample of bona-fide cluster members on the basis of their location on the vector-point diagram.
    \item[2. Global transformations:] for each star, we use the positions of all well-measured stars within a radius of 1000 NACO pixels to determine the optimal six-parameter transformation between the two master frames. \footnote{Six-parameter linear transformations also solve for the rotation between frames. By using cluster stars to compute the coefficients of such transformations, we are implicitly absorbing any potential systemic-rotation signal of the cluster and, as such, the rotation in the plane of the sky of NGC 6441 cannot be directly detected with our PMs (see the discussion in, e.g., \citealt{bellini_hubble_2017} and \citealt{libralato_hubble_2018}).}
    \item[3. Calculation of PM and PM error:] the stellar PMs are obtained as the difference between the transformed positions of the two master frames, divided by the temporal baseline. The PM errors are calculated by quadratically adding the positional errors of the single epochs and then dividing this result by the temporal baseline.
\end{description}

\subsection{A posteriori local corrections}

By construction, the mean motion of cluster members should be at location (0,0) on the vector-point diagram regardless of stellar positions on the master frame. Local deviations of the mean motion are caused by small, uncorrected-for systematic effects, which we mitigated using an a-posteriori local correction as follows.

For each cluster star in our PM catalogue, we chose the 50 closest neighbouring cluster stars and calculated their mean motion. This value is subtracted from the measured PM of the star. This step leads to an additional statistical PM error of $\sim$0.028~mas~yr$^{-1}$ per coordinate. Therefore, we kept both the uncorrected and the corrected position residuals for the further analysis.

\subsection{Resulting proper motion precision}

\begin{figure*}
  \centering
    \includegraphics[width=1.0\textwidth]{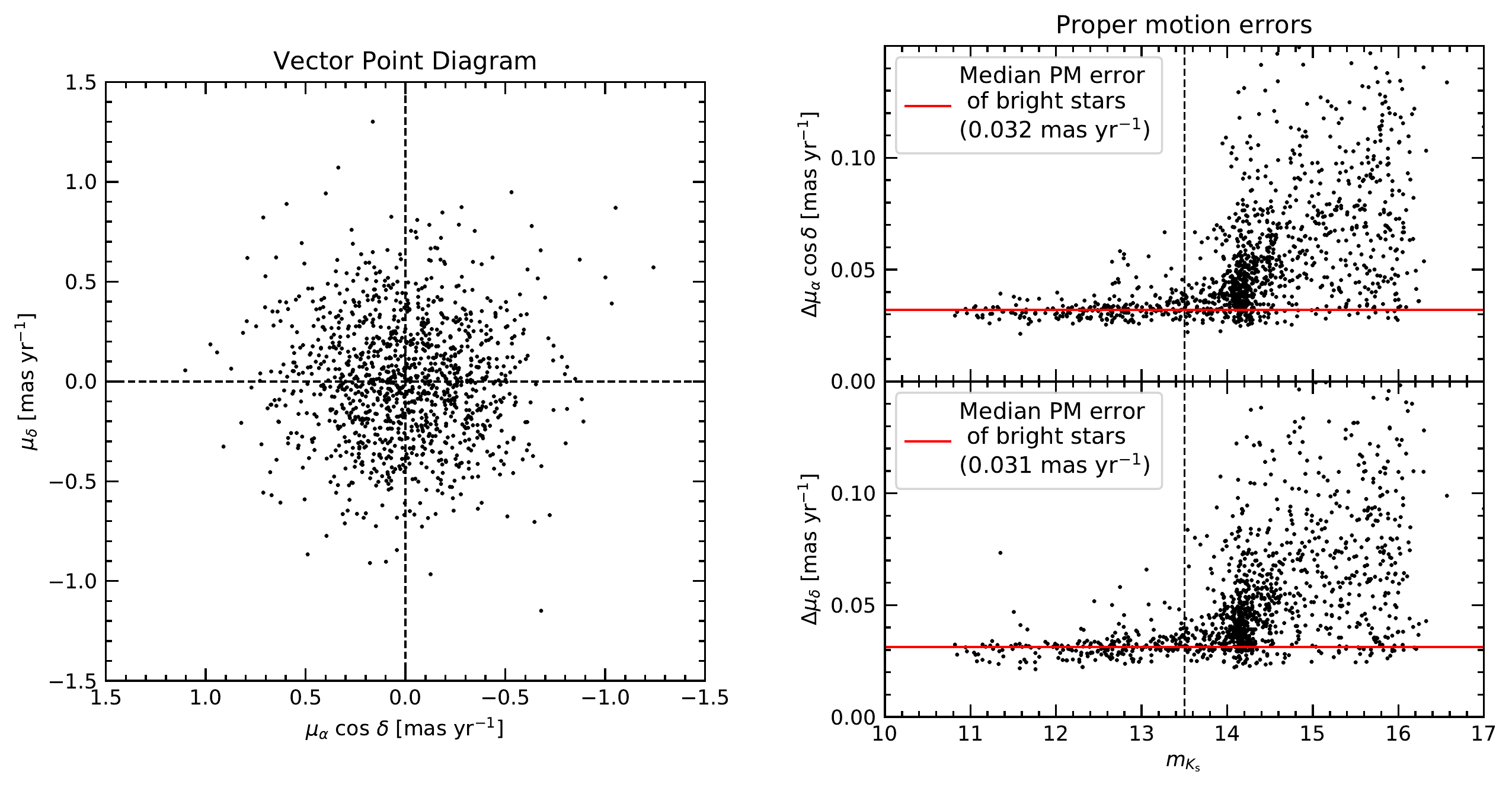}
      \caption{The left panel shows the vector-point diagram with our measured PMs for NGC~6441. The sample is restricted to stars that have been used for the kinematic analysis. The two right panels show the magnitude dependence of the PM errors for the stars in the sample. The red line marks the median PM error for stars brighter than $m_{\rm K_{\rm S}}=13.5$ ($\sim m_{\rm F555W}=17.5$). One can see that for these stars, a PM precision of around 0.03~mas~yr$^{-1}$ per PM component is reached.}
  \label{fig:vpd_pme}
    
\end{figure*}

The vector-point diagram and the PM uncertainties in both coordinates are shown in \autoref{fig:vpd_pme}. For bright stars with $m_{K_{\rm S}}<13.5$ (or $m_{\rm F555W} < 17.5$) we reach uncertainties of around 0.03~mas~yr$^{-1}$. Similar precisions are reached in pure \textit{HST} based studies \citep[see][]{bellini_hubble_2014} while the \textit{Gaia} DR2 PMs have a higher uncertainty of around 0.1~mas~yr$^{-1}$ for stars with G=17~mag even in less crowded fields \citep{lindegren_gaia_2018}.

\section{The kinematics of NGC~6441}
\label{sec:kin}
\subsection{Determination of the velocity dispersion}
\label{section:dispersion}
Only cluster stars (within 1.5~mas yr$^{-1}$ from the origin of the vector-point diagram) that are well-measured in the \textit{HST} and NACO data were used in the analysis of the internal kinematics. For the \textit{HST} data, we defined as ``well-measured'' stars that have: (i) magnitude rms in both filters lower than 0.025 mag; (ii) \texttt{QFIT} parameter larger than 0.985; (iii) the absolute value of the shape parameter \texttt{RADXS} lower than 0.03. For the NACO data, we only required the 1D positional errors to be lower than 0.4 NACO pixel. Stars with a PM error in either direction greater than 0.15~mas yr$^{-1}$ or larger than half the local velocity dispersion $\sigma_\mu$ of the closest 50 cluster stars were excluded from the analysis. We tested different quality selections, and find negligible differences in the resulting velocity dispersions. The parameters described above provide a good compromise between including as many stars as possible (for better statistics in the kinematic analysis) and excluding poorly-measured objects. Out of the around 1400 stars with a PM measurement, $\sim$1200 were included in the kinematic analysis.

The velocity dispersions were obtained by subtracting in quadrature the PM errors from the observed scatter of the PMs \citep{van_der_marel_new_2010}. We analysed the combined ($\sigma_\mu$), radial ($\sigma_{\rm R}$), and tangential ($\sigma_{\rm T}$) velocity dispersions as a function of distance from the cluster's centre in: i) one radial bin with all stars within 1 arcsec from the cluster's centre (23 stars); ii) four equally-populated radial bins between 1 and 5 arcsec (107 stars each); iii) eight radial bins using all stars outside the centermost 5 arcsec (seven groups of 97 stars and one with 87).

The results are shown in \autoref{fig:vd_results_with_comparison} and in Table \ref{data_table} in the Appendix. Our data are very consistent with the PM results of \cite{watkins_hubble_2015} and the MUSE LOS measurements of \cite{kamann_stellar_2018}, which are also shown in \autoref{fig:vd_results_with_comparison}. However, we cannot reproduce the dip in velocity dispersion in the innermost MUSE data point, that may be caused by crowding effects \citep{alfaro-cuello_deep_2020}. In the innermost arcsecond we measure a combined velocity dispersion of (0.316~$\pm$~0.034)~mas~yr$^{-1}$. If we employ the dynamical distance estimate from Section \ref{sec:kin} ($D = 12.74$~kpc) this corresponds to (19.1~$\pm$~2.0)~km~s$^{-1}$.
\begin{figure*}
  \centering
    \includegraphics[width=1.0\textwidth]{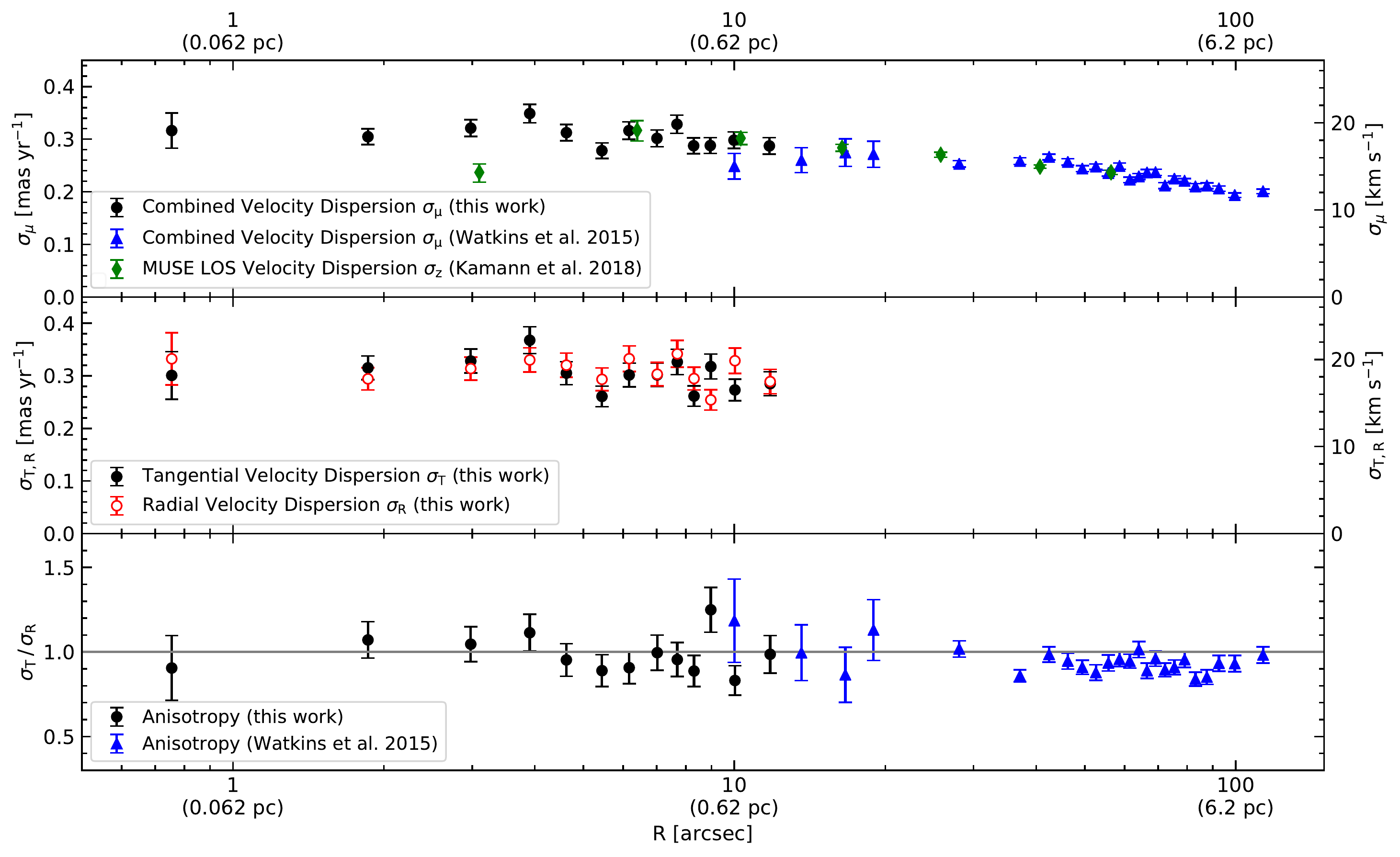}
      \caption{This plot shows the results of the determination of the velocity dispersion both in units of mas~yr$^{-1}$ and km~s$^{-1}$ (converted using our best estimate for the dynamical distance of 12.74~kpc). The upper panel shows the combined velocity dispersion (calculated by treating the tangential and radial proper motion as individual measurements). The middle panel shows the velocity dispersion calculated separately for the two components. The lower panel shows the ratio between the two, i.e., the anisotropy of the cluster.}
  \label{fig:vd_results_with_comparison}
\end{figure*}

\subsection{Search for fast-moving stars}
Fast moving stars in the centre of NGC~6441 could be a possible signature of an IMBH \citep{drukier_can_2003}. The lack of LOS velocities does not allow us to obtain a 3D kinematic picture of the core of this cluster. However, we can still investigate the presence and nature of fast-moving objects in the plane of the sky thanks to our high-precision PMs.

We initially searched for stars with a PM that indicates a velocity higher than the escape velocity of the cluster as a result of the interaction with a potential IMBH. We considered fast-moving stars those with total PM between 1.26 and 1.5~mas~yr$^{-1}$. The lower limit corresponds to the escape velocity of $v_{\rm esc} = $76~km~s$^{-1}$ reported by \cite{baumgardt_catalogue_2018} for NGC~6441. We find 4 stars in this velocity range. Stars with a PM~$>$~1.5~mas~yr$^{-1}$ are not considered in our analysis because they are either mismatches between the NACO and \textit{HST} catalogues, or are located in the outskirts of our field of view, far away from the cluster centre.

For each high-PM star, we investigated if its PM vector suggests a past passage within 2.5 arcsec from the centre of the cluster (for a similar procedure see \citealt{2021MNRAS.500.3213L}). This specific radius is the influence radius of an IMBH with a mass of $M_{\rm BH}=1.3\times10^4$~M$_\odot$, i.e.,  the upper limit of the IMBH mass we computed in Sect.~\ref{modelresults}. We find only one high-PM star with a PM vector consistent with an ejection caused by the interaction with an IMBH in the core of the cluster.

\begin{figure}
  \centering
    \includegraphics[width=0.5\textwidth]{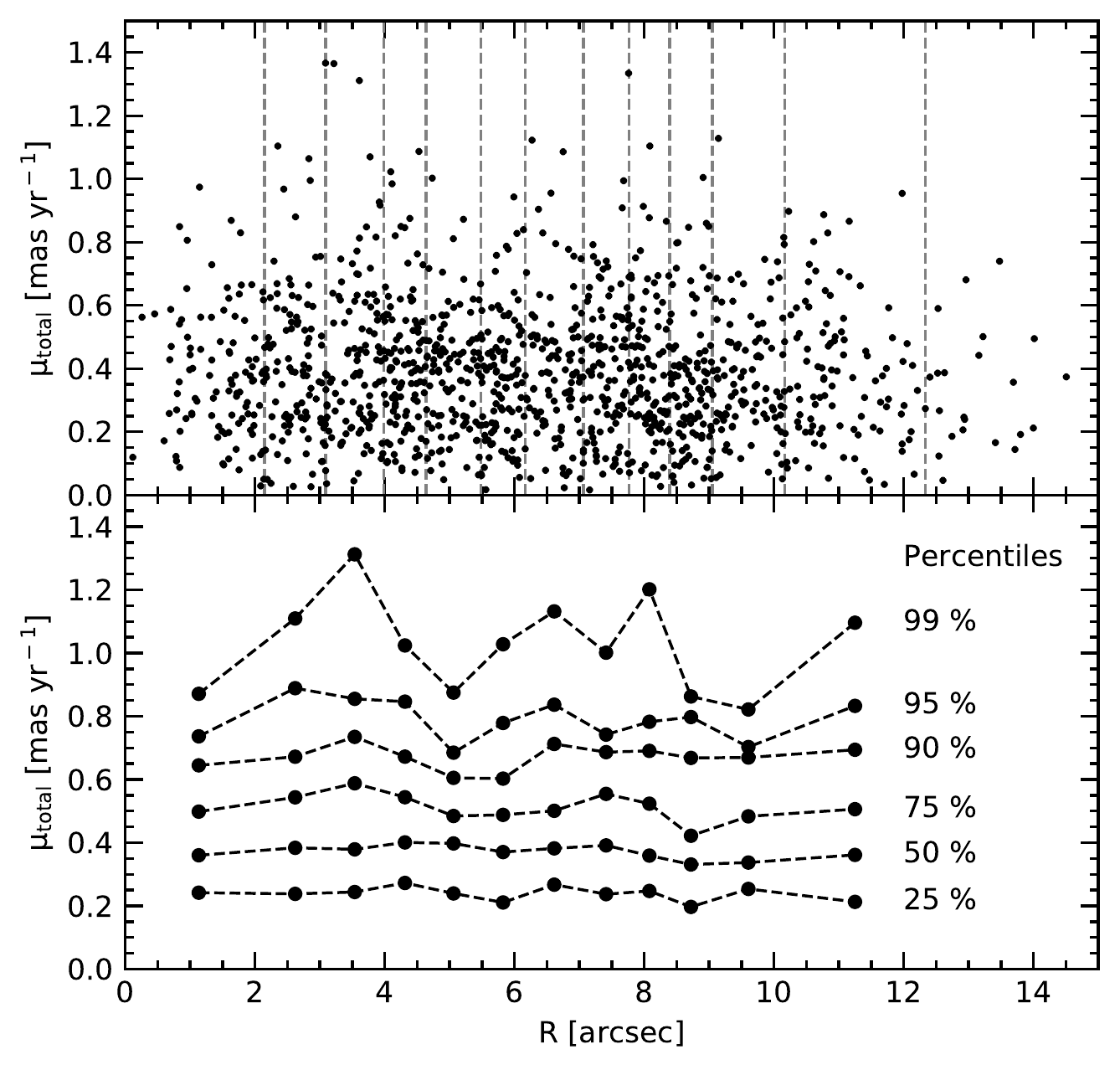}
      \caption{The upper panel shows the total proper motion (i.e. the quadratic sum of both components) of all stars in our measured sample and their distance from the centre. We separate the sample in radial bins of 100 stars each. The lower plot shows different percentiles of the PM in each bin.}
 \label{fig:fast_moving_stars}
\end{figure}

We also performed a statistical analysis of the total PMs as done by \citet{anderson_new_2010} for the globular cluster $\omega$~Centauri. We divided our sample in 12 radial bins of 100 stars each, and determined various percentiles of the total PM distribution in each such bin (see \autoref{fig:fast_moving_stars}). If an IMBH is harboured in the core of NGC~6441, we would expect a higher number of high-velocity stars closer to the centre. However, we do not detect a significant difference between the distribution within the innermost 2 arcseconds and the distributions in the outer bins at $R>8$~arcsec. 

We can conclude that, even though we do detect a single fast-moving star that is geometrically associated with the cluster centre, this is not enough to clearly indicate the presence of an IMBH. Deeper observations would be necessary to expand our brightness-limited sample and only the combination of radial velocity measurements with proper motions can give the full velocity vector.

\subsection{Dynamical Models}

Kinematics tell us how fast the stars are moving inside the cluster. Dynamical models offer us a way to determine the underlying physics that make the stars move as they do. We ran dynamical models of NGC~6441 to determine its mass profile, anisotropy profile, and dynamical distance. We also used our models to assess whether we could determine the existence and properties of a possible IMBH at the centre of the cluster.
\subsubsection{Setup of models}
We used a compilation of 4 kinematic datasets. We started with the NACO PM catalogue of well-measured stars that was used to determine the velocity dispersion profile of the cluster in Section \ref{section:dispersion}. We augmented this with the catalogue of \textit{HST} PMs from \citet{Watkins2015b}, which probe further out in the cluster than the NACO dataset alone. To this, we added the LOS velocity dispersion profiles from \citet{kamann_stellar_2018} and \citet{baumgardt_catalogue_2018}. We used the surface brightness profile from \citet{Trager1995} as a proxy for the surface number density profile from which the kinematic samples were drawn.

For the dynamical models themselves, we used the spherical Jeans Anisotropic Multi-Gaussian-Expansion (JAM, MGE) models \citep{Cappellari2008, Cappellari2015}. The methodology broadly follows that described in Section 5.2 and Appendix C of \citet{HenaultBrunet2019}, and we refer to that paper for details. Here we briefly summarise the analysis.

In the JAM models, the stellar and mass densities are treated as MGEs. Each component of the MGEs has a width and a weight. We assumed that the widths were the same for the stellar and mass MGEs but allowed the weights to vary independently so as to fit a variable mass-to-light ratio (M/L). By giving each component of the stellar MGE an anisotropy, we were also able to fit for a variable anisotropy profile. We used 5 Gaussian components to fit the cluster. We tried initially with 6 and found that this was too many, so reduced to 5 and found that this gave much improved and non-degenerate results. Altogether, this gave us 20 free parameters: 5 widths $s_i$, 5 stellar density weights $\nu_i$, 5 mass density weights $\rho_i$, and 5 anisotropies $\beta'_i$. There were two further parameters -- the distance $D$, and the IMBH mass $M_\mathrm{BH}$ -- taking the total number of free parameters to 22.

In the spherical JAM models, the mean velocities are everywhere zero, and the velocity distributions are characterised by the dispersions in the projected-radial, projected-tangential, and LOS directions. The likelihood of the data given the model was calculated differently for the PMs and the LOS velocities due to the different datasets available. We treated the PMs discretely, that is we did not bin the stars, but instead for each star calculated the likelihood of observing a star with the measured velocity and uncertainty given the mean velocity and velocity dispersion predicted by the model at the position of the star, assuming Gaussian velocity distributions and uncertainties. For the LOS velocities, we had only binned velocity dispersion profiles, so we calculated the likelihood of measuring a given velocity dispersion and uncertainty given the velocity dispersion predicted by the model at the position of the bin.

The Gaussian widths, Gaussian weights and the IMBH mass were all fitted in log-space, as they had the potential to span many orders of magnitude. We used priors to restrict the range of certain parameters for both physical and computational reasons, but otherwise used flat priors within the allowed ranges.

To avoid degeneracies between different components and to ensure that the models fit for 5 separate components, we insisted that $\log s_{i+1} - \log s_i >0.2$. Additionally, we limited the widths of the innermost and outermost MGE components such that $s_1 > R_{25}$ and $s_5 < {R_\mathrm{max}/\sqrt{3}}$, where $R_{25}$ and $R_\mathrm{max}$ are the projected radial positions of the 25-th star in the PM dataset and outermost point in surface brightness profile, respectively. The first condition ensured that there were at least 25 stars inside of $s_1$ to constrain the inner cluster properties, and the second condition ensured that the outer slope of the density profiles was at least 3, which is required for a finite system.

The light and mass weights were assumed to be positive. We further added a lower limit on the M/L of each component such that $\rho_i/\nu_i > 0.1$, as values lower than this would be unrealistic physically.

$\beta'$ is a modified anisotropy
\begin{equation} 
    \beta' =  \frac{\sigma_r^2 - \sigma_t^2}{\sigma_r^2 + \sigma_t^2}
    \label{eqn:betaprime}
\end{equation}
where $\sigma_r$ and $\sigma_t$ are the velocity dispersions in the radial and tangential directions in a spherical coordinate system. This modified anisotropy is defined to be both symmetric about isotropy ($\beta'=0$) and finite in extent ($\beta' = \pm 1$). In practice, we only sample between $\beta' \approx \pm 0.88$\footnote{The actual limits correspond to cases where one dispersion is 4 or $\frac{1}{4}$ times the other.} to avoid extreme anisotropies that are computationally intensive but not seen in real clusters.

Finally, we restricted how much the density and anisotropy profiles could change from component to component by insisting that $-7 < \Delta_{\rho,\nu} / \Delta_s < 3$ and $-2 < \Delta_{\beta'} / \Delta_s < 2$, where $\Delta_\rho = \rho_{i+1} - \rho_i$ and similarly for $\Delta_\nu$ and $\Delta_s$ and $\Delta_{\beta'}$. In practice, very few components hit the limits of these ranges.

We set a lower limit of the central black hole mass of 0.1~\Msun; the BH mass was fit in log space and very small masses are indistinguishable in the model so this parameter has the potential to go to $-\infty$ if not limited. No other restrictions were set. The distance was fit in linear space, and was assumed to be positive with no further restrictions.
\subsubsection{Modelling results}\label{modelresults}
\begin{figure*}
    \centering
    \includegraphics[width=\linewidth]{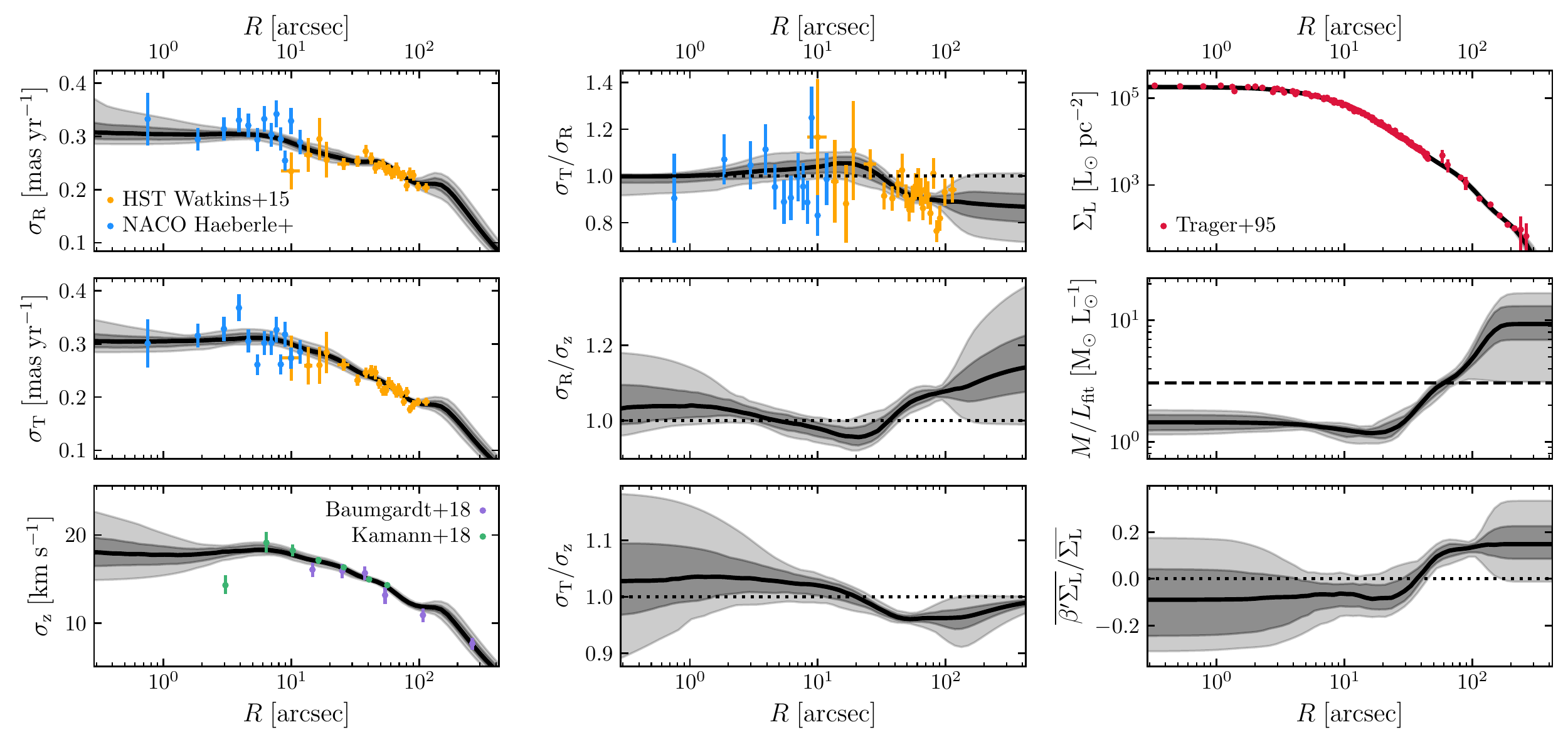}
    \caption{Best-fitting model profiles. From left to right then top to bottom: radial PM dispersion, tangential PM dispersion, LOS velocity dispersion, tangential PM / radial PM anisotropy, radial PM / LOS velocity anisotropy, tangential PM / LOS velocity anisotropy; surface brightness, projected M/L, proxy for projected spherical anisotropy. In all panels, the solid black line is the median of the fits, the darker shaded regions span the 15.9 to 84.1 percentiles (approximately the 1-sigma confidence region) and the lighter shaded regions span the 2.5 to 97.5 percentiles (approximately the 2-sigma confidence region). The coloured data points show data from this work and from literature sources to which the fits were performed as shown. In the anisotropy panels, the dotted lines highlight isotropy. In the mass-to-light panel, the dashed line shows the global M/L for the cluster calculated as the total mass over the total light. Overall, the fits to the data points are good. The cluster is best fit by a variable M/L, consistent with a cluster with some mass segregation, and a variable anisotropy profile, consistent with a cluster that is more relaxed in the central regions than the outer parts due to their different relaxation times.}
    \label{figure:ProfileFits}
\end{figure*}

We explored the available parameter space using the affine-invariant Markov Chain Monte Carlo software \textsc{emcee} \citep{ForemanMackey2013}. We used 100 walkers and ran for 7\,500 steps to be sure that the chains had fully converged. The resulting fits to the cluster are shown in \autoref{figure:ProfileFits}. In each panel, the coloured data points show the data with their uncertainties (in cases where there are no points, data is either not available or not measurable). The black solid lines show the median of the fits for the profile, the darker shaded region spans the 15.9 to 84.1 percentiles (equivalent to the 1-sigma confidence region for a Gaussian distribution) and the light shaded region spans the 2.5 to 97.5 percentiles (equivalent to the 2-sigma confidence region for a Gaussian distribution).

The left panels show from top to bottom the projected radial dispersion profile $\sigma_{\rm R}$, the projected tangential dispersion profile $\sigma_{\rm T}$, and the LOS velocity dispersion profile $\sigma_{\rm z}$. In the top and middle panels, the blue points show the profiles calculated from NACO PMs and the orange points show the profiles calculated from the \textit{HST} PMs, however we stress that the fits were not done to these profiles but to the individual measurements. The profiles are shown here for visualisation purposes. In the bottom panel, the green points show the MUSE dispersion profile from \citet{kamann_stellar_2018} and the purple points the dispersion profile from \citet{baumgardt_catalogue_2018}; for these we did fit directly to the binned dispersion profiles. In general, these fits are very good. The models struggle in the outer regions where there are only LOS constraints but not PMs, and show some broadening in the centre where there are very few stars. The only poorly-fit point is the central point in the MUSE LOS dispersion profile, which the models were not able to constrain. \citet{kamann_stellar_2018} also point out this central dip in their dispersion profile; they note that it could potentially be due to crowding but argue that this is likely not the case as the cluster does not have a steep central surface brightness profile. That the models are unable to reconcile the PM and LOS dispersion profile may suggest otherwise and that the central MUSE data point is more affected by crowding than previously believed. This effect due to crowding is also clearly seen in the MUSE velocity dispersion profile of the cluster M54 \citep{alfaro-cuello_deep_2019, alfaro-cuello_deep_2020} and can be overcome by higher spatial resolution data (Alfaro-Cuello et al. in prep.).

The panels in the central column show the anisotropies, from top to bottom they are tangential over radial, radial PM over LOS, and tangential PM over LOS. For the top panel, we are able to show data (NACO PMs in blue and \textit{HST} PMs in orange as for the PM dispersion panels) as we have the same set of bins for both the radial PM and tangential PM datasets. For the middle and lower panels, we show only the model fits, where we have used the distance of each model to convert the PMs from mas~yr$^{-1}$ to km~s$^{-1}$ so that isotropy is at 1; as we do not have the same data coverage (and, hence, bins) for the LOS and PM samples we cannot calculate these anisotropies directly, but the models given us an insight into what we cannot measure easily from the existing data. All panels are not constant, indicating that there is some anisotropy in the system and it varies through the cluster.

The right column shows, from top to bottom, the surface brightness profile, the projected M/L, and a proxy for the projected spherical anisotropy profile. The surface brightness profile is a key part of the model and so we are able to compare the model fit to the data, and it is a very good fit overall.

We cannot measure the mass directly -- indeed, this is one of the main motivators for carrying out these models -- so the M/L profile has no data points against which to compare. The M/L is clearly not constant through the cluster. It is $\sim1.5\,\text{M}_\odot/\text{L}_\odot$ at the centre, falls slightly in the intermediate regions and then increases to $\sim10\,\text{M}_\odot/\text{L}_\odot$ in the outer parts. This is consistent with a cluster that has some mass segregation, whereby high-mass stars tend to be more centrally concentrated than low-mass stars. The difference in mass between high-mass stars and low-mass stars is only around a factor of a few, 10 at most, but the low-mass stars are far more numerous than the high-mass stars, so it is the low-mass stars that dominate the mass budget. However, the high-mass stars are many hundreds or thousands times brighter than the low-mass stars, so it is the high-mass stars that dominate the light budget, despite their lower numbers. This interplay typically gives an M/L profile much like what we see here. The dashed line shows the global M/L of $\sim$3.0 for the cluster calculated as the total mass divided by the total light, which is consistent with stellar population synthesis estimates for clusters of its metallicity. This over-predicts the true M/L in the inner regions and under-predicts the true M/L in the outer regions.

Our best estimate for the total mass of the cluster is $2.38^{+0.37}_{-0.30}\times 10^6\,\text{M}_\odot$. This value is higher than the values reported by other authors. \cite{mclaughlin_resolved_2005} determined a mass in the range of  ($1.45^{+0.28}_{-0.25} - 1.86^{+0.33}_{-0.28})\times 10^6\,\text{M}_\odot$ based on M/L values of stellar population fits. \cite{baumgardt_catalogue_2018} report a mass of \num{1.2e6} $\text{M}_\odot$ based on a comparison between isotropic N-body models of globular clusters with observed LOS velocity measurements. We advise caution in interpreting these global measurements. The profiles are best constrained where there is a lot of kinematic data (the intermediate radii) and less well constrained where the data is sparse (the innermost and outermost regions). In particular, the outer regions of the mass profile are driven by the outermost datapoint in the \citep{baumgardt_catalogue_2018} LOS profile, where there is no corresponding PM data at all. This is reflected in the large scatter we see in the outer regions of the fits.

The middle panels of \autoref{figure:ProfileFits} show how the dispersions measured in two orthogonal directions compare in a 3-dimensional system. The lower panel in the right column effectively shows the 3-dimensional (spherical) anisotropy. Recall, in the JAM models, each component of the stellar density MGE is assigned an anisotropy $\beta'$. We calculate a luminosity-weighted anisotropy profile by multiplying the stellar density weights by the anisotropy and calculating the corresponding MGE profile. We divide this by the stellar density (surface brightness) MGE to get a proxy for the anisotropy profile. Here isotropy is at 0. The slight negative bias at the centre indicates some mild (spherical) tangential anisotropy there, although the models are also consistent with isotropy. The positive bias in the outer parts indicates (spherical) radial anisotropy there. This trend is consistent with theoretical expectations for cluster evolution and with cluster simulations. Clusters are expected to develop with some radial anisotropy, but they become more isotropic as they relax \citep{baumgardt_dynamical_2003, vesperini_kinematical_2014, tiongco_velocity_2016}. As relaxation times are shorter at the centres, the centres relax and reach isotropy first. For NGC~6441, \cite{harris_new_2010} reports a half-mass relaxation time of $\textrm{log}(t_{\textrm{hm}}/\rm yr)=9.09$, while the core relaxation time is significantly shorter with $\textrm{log}(t_{\textrm{core}}/\rm yr)=7.93$.

In addition, we made use of the mass profile determined with our model to recalculate the relaxation times. To do so, we used the equations from \cite{1993ASPC...50..373D} with the modification of \cite{harris_new_2010}. We took the core radius value of $r_{\rm core}=0.13$~arcsec from \cite{harris_new_2010} (converted to $r_{\rm core}=0.48$~pc using our dynamical distance estimate),  and assumed an average stellar mass of $\frac{1}{3}$~M$_\odot$. The other parameters are based on our model. With our total mass estimate of $2.38^{+0.37}_{-0.30}\times 10^6\,\text{M}_\odot$, we find a half-mass radius of $8.0^{+1.0}_{-0.7}$~pc. We determine a core density of $(1.35^{+0.19}_{-0.13})\times 10^5$~M$_\odot$~pc$^{-3}$. The resulting values for the relaxation times are $\textrm{log}(t_{\textrm{hm}}/\rm yr)=10.16^{+0.11}_{-0.08}$ in the core, and $\textrm{log}(t_{\textrm{core}}/\rm yr)=7.84^{+0.02}_{-0.02}$ at the half-mass radius.

While our values for the relaxation time in the core are similar to the values in the literature, our value for the half-mass relaxation time is larger as the previously reported values mainly because of the larger estimate for the half-mass radius. Due to the variable M/L ratio and our high total mass estimate, this value is higher then the half-light radius, which was used as a proxy for the half-mass radius in previous studies.

Although we advise caution in interpreting these global values (see above), we can see that the relaxation time at the half-mass radius is longer than in the core. This difference in relaxation times could explain why we observe isotropy in the centre, but anisotropy in the outer regions of the cluster. Similar astrometric findings for other clusters are discussed in \cite{watkins_hubble_2015}.

\begin{figure}
    \centering
    \includegraphics[width=\linewidth]{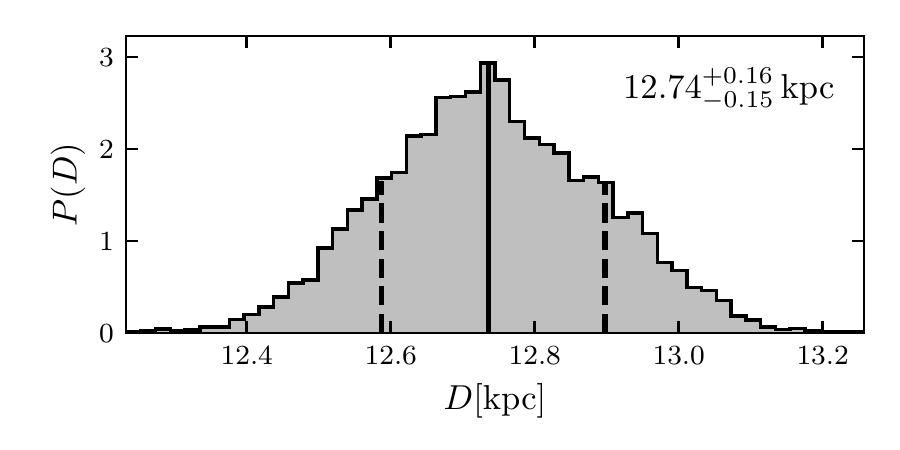}
    \includegraphics[width=\linewidth]{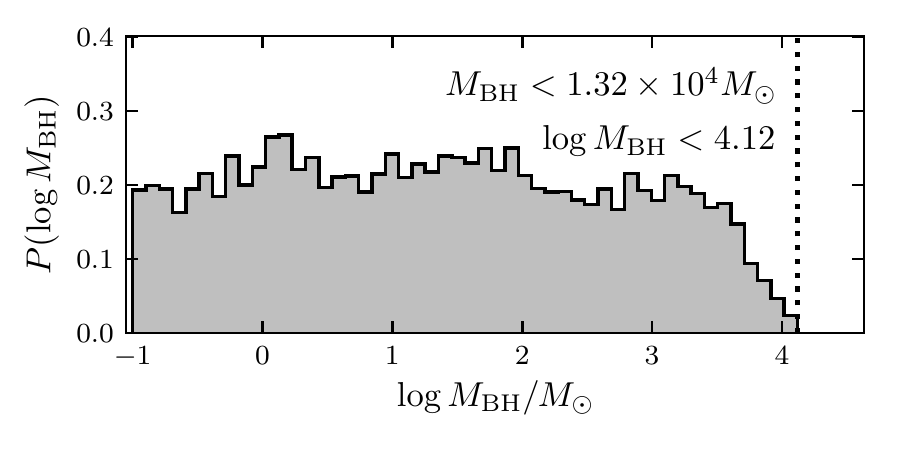}
    \caption{The final distributions of the distance (top) and IMBH mass (bottom) parameters at the end of the MCMC chains. The distance is well constrained and the median and 15.9 and 84.1 percentiles were used to provide a distance estimate as shown in the top right corner. The IMBH mass is unconstrained and we can neither confirm nor rule out the presence of an IMBH, but we are able to place an upper limit on a possible IMBH mass, as given in the top right corner.}
    \label{figure:histograms}
\end{figure}

\autoref{figure:histograms} shows the distribution of distance (top panel) and IMBH mass (bottom panel) estimates at the end of the MCMC run. To construct these plots, we selected 10\,000 points in total, 100 walkers from each of 100 steps at 50 step intervals. The distance is nicely constrained, and approximately Gaussian overall. To determine a distance estimate and uncertainty, we calculate the median and 15.9 and 84.1 percentiles of the distribution to obtain distance $D = 12.74 ^{+0.16} _{-0.15}$~kpc. This estimate is in reasonable agreement with the estimate from \citet{kamann_stellar_2018} of $12.5 \pm 1.2$~kpc but offset from the values of 11.6~kpc from \citet{harris_new_2010} and of $11.83 \pm 0.14$~kpc from \cite{baumgardt_mean_2019}.

The putative IMBH mass is not at all well constrained as evidenced by the flat distribution in the lower panel of \autoref{figure:histograms}. With this data we cannot conclusively state whether or not an IMBH is present as models with a fairly massive BH and models with sub-solar-mass (effectively no) BH are equally likely, although models with very massive BHs are ruled out. The best we can do here is a place an upper limit on the mass of a possible IMBH at $M_\mathrm{BH} < 1.32 \times 10^4$~\Msun.
\subsubsection{Context of our IMBH mass limit}
There are several scaling relations for black-hole masses in galactic bulges that are valid over a wide mass range. We extrapolate them to the mass of NGC~6441 and compare the predicted black hole mass with our upper limit.

Using our estimate for the mass of the cluster of $2.38^{+0.37}_{-0.30}\times 10^6\,\text{M}_\odot$, the extrapolation of the relation between the Bulge and BH masses of \cite{schutte_black_2019} gives an expected black hole mass of $1.17^{+0.23}_{-0.18}\times 10^3\,\text{M}_\odot$, significantly smaller than our upper limit.

If we use the relation between velocity dispersion and BH mass of \cite{gebhardt_relationship_2000}, our central velocity dispersion of (19.1~$\pm$~2.0)~km~s$^{-1}$ translates in an IMBH mass of $1.8^{+0.37}_{-0.30}\times 10^4\,\text{M}_\odot$ which is quite similar to our upper limit.

Although these estimates are consistent with our upper limit, they are extrapolated from measurements of black-hole masses in a different (higher) mass regime than that of IMBHs in GCs. A more similar mass range is found in nuclear star clusters, in which the mass ratio between stellar cluster mass and black hole is typically $M_\textrm{BH}/M_\textrm{NSC}\approx 0.25$ with a scatter of 2 \citep{nguyen_nearby_2018, greene_intermediate-mass_2019}. This is much higher than the upper limit of $M_\textrm{BH}/M_\textrm{NGC\,6441}\leq 0.0054$ we obtain for NGC~6441.

Theoretical predictions for IMBH masses in globular clusters have quite a range range between 0.001 and 0.01 of the cluster mass \citep{miller_production_2002, portegies_zwart_runaway_2002, giersz_mocca_2015}, which is in agreement with our measured upper limit.

Finally \cite{tremou_maveric_2018} report an upper limit of $\textrm{M}_\textrm{BH}<2270\,\text{M}_\odot$, based on the (missing) radio signature of an accreting IMBH in NGC~6441.

We can conclude that a larger sample of stars with precise PMs in the central region is necessary to observationally reach the lower end of the predicted black hole mass range and to put a final answer on the existence/non-existence of an IMBH in this cluster.

\section{Conclusions}
\label{sec:con}

We determined PMs of around 1400 stars in the inner 15 arcseconds of the globular cluster NGC~6441 by combining space-based and ground-based position measurements taken 15 years apart. Because of the high astrometric precision of both epochs and the long timebase between them, our PM measurements reach a precision of 30~\textmu as~yr$^{-1}$ for bright stars ($m_{K_{\rm S}} < 13.5$ or $m_\text{F555W} < 17.5$). Similar precisions are reached in pure \textit{HST}-based studies (see \citealt{bellini_hubble_2014}) while the \textit{Gaia} DR2 PMs have a higher uncertainty of around 0.1~mas~yr$^{-1}$ for stars with $G = 17$~mag in less crowded fields \citep{lindegren_gaia_2018}. This proves the potential of combined ground-based AO and space-based astrometry with a long temporal baseline.

With the PM data we were able to determine the velocity dispersion profile of evolved stars in core of the cluster. In the innermost arcsecond we measure a velocity dispersion of (0.316~$\pm$~0.034)~mas~yr$^{-1}$ which corresponds to (19.1~$\pm$~2.0)~km~s$^{-1}$ assuming a distance of 12.74~kpc.

Using our PM measurements, we searched for signatures of a potential IMBH. Although we find one fast-moving object, whose projected trajectory is compatible with being ejected from the core because of the interaction with an IMBH, a statistical analysis of the PMs in our field does not show any signs of the presence of an IMBH in the centre of NGC~6441. A complete 5D picture (including radial velocity measurements) and deeper observations of the core are still needed to clearly confirm or rule out any hypothesis.

We used Jeans models to fit a combination of our newly-obtained kinematic data and existing kinematic catalogues. From the best-fit models we could determine the underlying physical properties of the cluster: the global M/L of the cluster is $\sim3.0\,\text{M}_\odot/\text{L}_\odot$, but it varies from $\sim1.5\,\text{M}_\odot/\text{L}_\odot$ in the core of the cluster to $\sim10.0\,\text{M}_\odot/\text{L}_\odot$ in the outskirts. This is consistent with mass segregation where bright, high-mass stars are more centrally concentrated than low-mass stars.
In the core we do not observe significant anisotropy, however the outer parts of the cluster show some radial anisotropy.
By combining LOS velocity measurements with our PM measurements we obtain a dynamical distance of $D = 12.74 ^{+0.16} _{-0.15}$~kpc.
The models include a possible IMBH in the centre of the cluster. Our results are compatible with both the existence and non-existence of such a black hole, and we can only place an upper limit of  $M_\mathrm{BH} < 1.32 \times 10^4$~\Msun~on the mass of the black hole. This value is about one order of magnitude larger than the mass predicted by extrapolating the relation between BH and bulge masses, but consistent with other predictions for GCs.

In future studies, deeper observations of the cluster would be beneficial as the number of stars in our kinematic sample is clearly limited by the number of detectable stars in the NACO exposures. 

In the second half of this decade, instruments at extremely large telescopes, such as ELT MICADO, will allow PM measurements with an even higher precision and be able to finally answer the question of whether IMBHs are present in the cores of globular clusters.

\section*{Acknowledgements}

Based on observations with the NASA/ESA \textit{Hubble Space Telescope}, obtained at the Space Telescope Science Institute,which is operated by AURA, Inc., under NASA contract NAS5-26555.

Based on observations collected at the European Southern Observatory under ESO programme 0101.D-0385.

This work has made use of data from the European Space Agency (ESA) mission \textit{Gaia} (\url{https://www.cosmos.esa.int/gaia}), processed by the \textit{Gaia} Data Processing and Analysis Consortium (DPAC, \url{https://www.cosmos.esa.int/web/gaia/dpac/consortium}). Funding for the DPAC has been provided by national institutions, in particular the institutions participating in the \textit{Gaia} Multilateral Agreement.

MH acknowledges financial funding from the \textit{Studienstiftung des deutschen Volkes} that supported his visit at STScI.

Support for this work was provided by grants for \textit{HST} programs AR-14322 and AR-15055 provided by the Space Telescope Science Institute, which is operated by AURA, Inc., under NASA contract NAS 5-26555.

\section*{Data Availability}
The data underlying this article were accessed from the ESO Science Archive\footnote{\url{http://archive.eso.org}} and the Mikulski Archive for Space Telescopes (MAST)\footnote{\url{https://archive.stsci.edu/}}.
The numerical values of the velocity dispersion determined in this work are provided in the article (Table \ref{data_table}) and its online supplementary material.



\bibliographystyle{mnras}
\bibliography{references} 




\appendix
\section{Geometric distortion correction}
\label{app}
A set of geometric distortion corrections (GDCs) for NACO has already been published by \cite{plewa_pinpointing_2015,plewa_optical_2018}. However, the authors themselves noted that the distortion of the NACO detector is not stable over time, but shows abrupt changes most likely linked to instrument interventions. Furthermore, we do not know the exact definition of the PSF centre used to determine the literature GDC and there is a degeneracy between the PSF definition and the GDC. For these reasons, we decided to independently solve for the geometric distortion (GD) of the NACO detector using our data in order to achieve the best astrometric precision possible.

To determine the GDC, we tried both an autocalibration approach and the use of the ACS/HRC catalogue as a distortion-free reference.  For the scientific analysis, we relied on the calibration based on the external ACS/HRC catalogue, as the resulting geometric distortion model led to a much better correction with fewer residual distortions.

\subsection{Determination of the GDC using an external reference catalogue}
\subsubsection{The reference catalogue}
The \textit{HST} ACS/HRC observations of NGC~6441 have a very high precision and a very reliable distortion correction that reaches an accuracy < 0.01 HRC pixel \citep[see the Instrument Science Report][]{2004acs..rept...15A}. In comparison with the uncorrected NACO catalogues, they can be considered effectively distortion free. However, we have to take into account the 15-year long time baseline between the \textit{HST} and the NACO observations. The velocity dispersion in the cluster centre of around 18~km~s$^{-1}$ at a distance of 12.74~kpc (our dynamical distance estimate) leads to an rms displacement of around 0.4~NACO~pixel. While this effect is purely statistical and is averaged out when measurements of multiple stars are combined, it still leads to a decreased precision of the GDC and can mask smaller GD effects.

To overcome the limitations caused by the stellar motions, we made use of a PM catalogue for the core of NGC~6441 created using the HRC exposures from 2003 and WFC3/UVIS observations from 2014 and that will be the subject of a future paper (Bellini et al., in preparation). Suffice here to say that the data reduction and proper-motion computation of this catalogue closely followed the prescriptions given in great details in \cite{bellini_hubble_2014} and \cite{bellini_hst_2018}. The number of well-measured stars in the core of the cluster is relatively small due to the larger pixel scale of the WFC3/UVIS channel (40~mas~pixel$^{-1}$). On the other hand, we only use bright, well-measured stars to determine the GDC anyway.

The PMs in this catalogue were used to propagate the HRC 2003 positions to the epoch of the NACO observations. We only included stars in the reference catalogue with total PM error $\sigma_\text{PM} < 0.07$~mas~yr$^{-1}$, which corresponds to an error in the displacement of 0.08~NACO~pixel. Furthermore, we restricted the selection of reference stars using different quality criteria. In the end, around 1600 stars were available as reference stars.

\subsubsection{Matching individual frames on reference catalogue}
\label{GDC:Step1}
As a first step, we matched the astrometric catalogues containing uncorrected NACO positions onto the reference catalogue using linear transformations. The parameters of the linear transformations were determined using a least-squares fit. The cut-off radius for matching stars was initially 3.5 NACO pixels and it was progressively decreased down to 1 NACO pixel during the iterative process.

To avoid absorbing the linear distortion terms (the so-called skew) already at this level, we used 4-parameter transformations, which contain a shift in x and y direction, a rotation and a change of scale.

\begin{figure*}
\begin{subfigure}{.495\textwidth}
\label{fig:sub-first}
\centering
\includegraphics[width=0.9\textwidth]{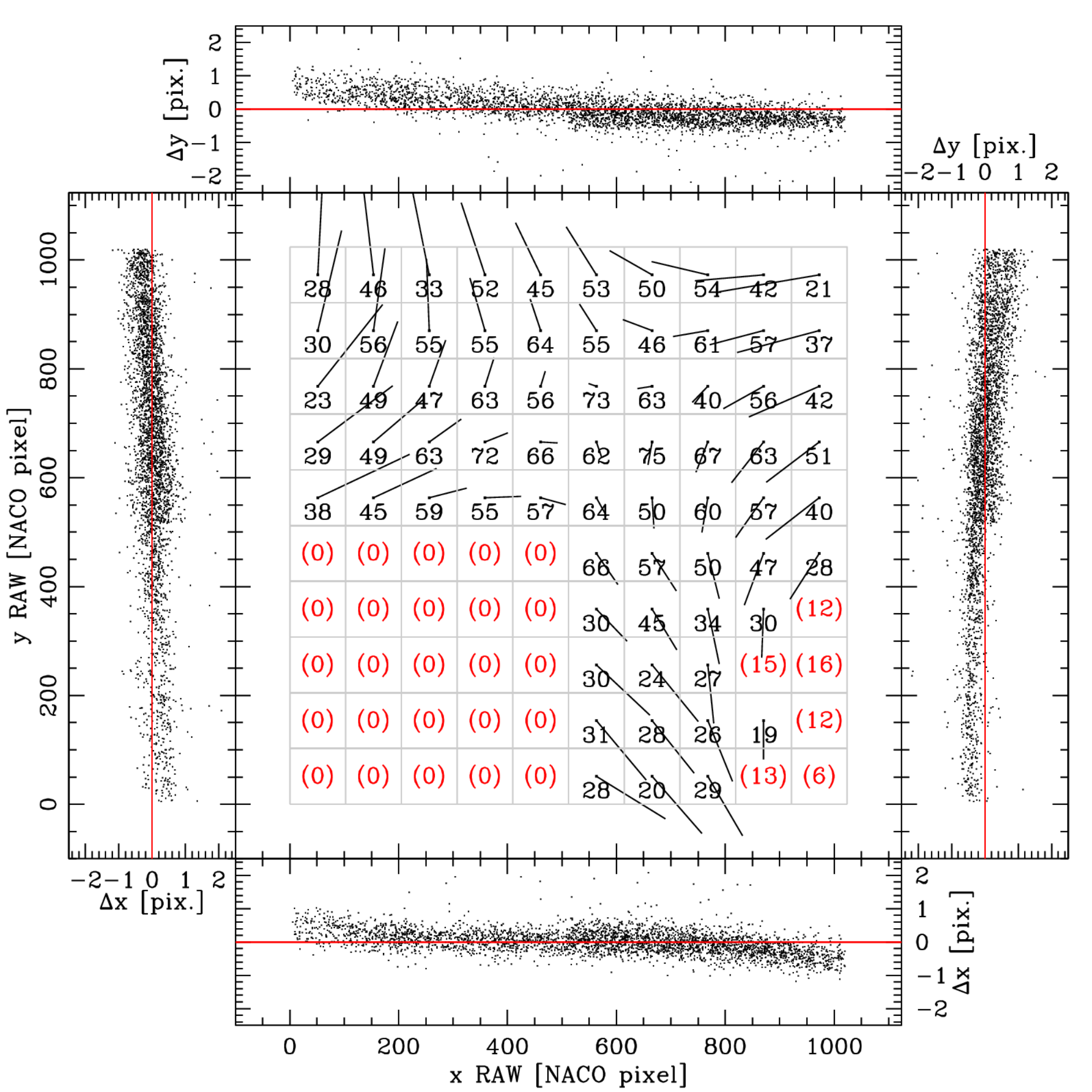}
\caption{Distortion map before the correction.\\rms of residuals: 0.321 NACO pixel}
\end{subfigure}
\begin{subfigure}{.495\textwidth}
\centering
\label{fig:sub-second}
\includegraphics[width=0.9\textwidth]{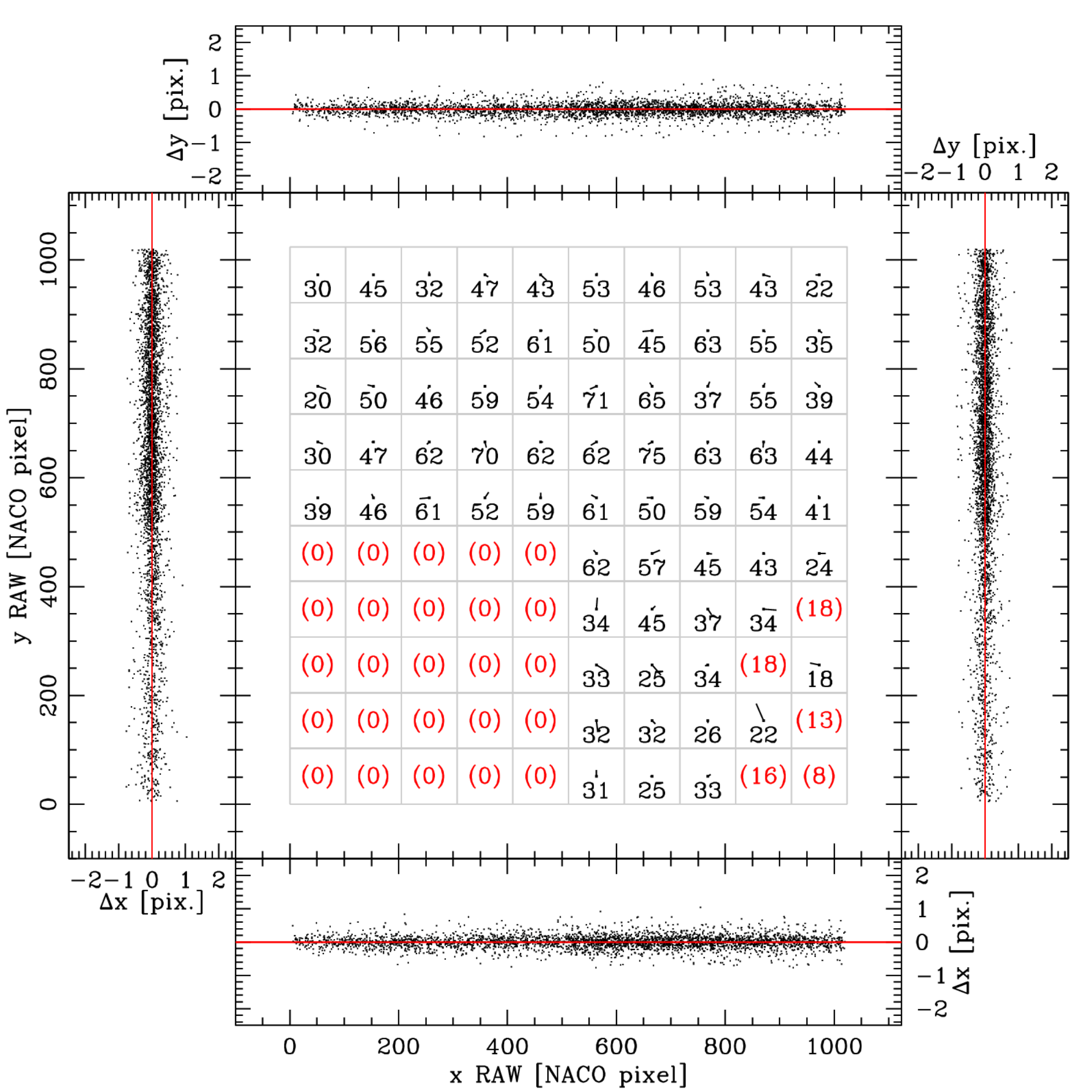} 
\caption{Distortion map after the correction.\\rms of residuals: 0.032 NACO pixel}
\end{subfigure}
\caption{The panels (a) and (b) show the distortion map of the NACO detector before (a) and after (b) the polynomial geometric distortion correction has been applied. The large central field shows a map of the detector with the grid of $10\times10$ binned residuals between the NACO measurements and the \textit{HST} reference. The distortion vectors are only plotted when more than 20 measurements fell in the respective bin. The length of the vectors is enlarged by a factor of 250. The side plots show all combinations of $x$ and $y$ residuals as a function $x$ and $y$ positions.}
\label{fig:gdc}
\end{figure*}

The residuals of the linear-transformation fit now contained both the individual measurement errors, and also the geometric distortion of the NACO images. We divided the detector in $10\times10$ quadratic bins and collected the residuals from all matched NACO catalogues. We then calculated the 3$\sigma$-clipped median of the residuals in each bin containing at least 20 residuals. The maps of the binned residuals can be seen in \autoref{fig:gdc} and provide a first indicator of how the geometric distortion correction will look like.
\subsubsection{Fit of a 2D polynomial model}
\label{GDC:Step2}
We modelled the binned residuals using the following 2D third order standard polynomials:
\begin{equation}
\begin{aligned}
     \delta x ={} &a_1 x + a_2 y\\
     + &a_3 x^2 + a_4 x y + a_5 y^2\\
     + &a_6 x^3 + a_7 x^2 y + a_8 x y^2 + a_9 y^3\\
     \delta y ={} &b_1 x + b_2 y\\
     + &b_3 x^2 + b_4 x y + b_5 y^2\\
     + &b_6 x^3 + b_7 x^2 y + b_8 x y^2 + b_9 y^3\\
\end{aligned}
\end{equation}
We used a least-square fit to determine the coefficients. The coefficients $a_1$ and $a_2$ were set to zero to lower the degrees of freedom of the fit and to enforce that, at the centre of the detector, our GDC will lead to the same $x$-scale as the detector and the corrected and raw $y$-axis will be aligned. However, as we cannot assume that $x$ and $y$ axis have the same scale/are perpendicular, we can not set the parameters $b_1$ and $b_2$ to zero.

A higher order of the polynomials did not lead to a better fit of the data. Also, the use of a different polynomial base (e.g.~Zernike polynomials) had no significant influence on the result.
\subsubsection{Iterative Process}
To avoid that the linear transformations between the single NACO catalogues and the reference frame are biased by uncorrected geometric distortion, we repeated the procedures described in an iterative process. After the polynomial coefficients have been determined, we applied 75\% of the corrections to the raw NACO catalogues and repeated the full process (described in \ref{GDC:Step1} and \ref{GDC:Step2}) with the new corrected coordinates until the polynomial coefficients converge (in our case after 200 iterations).
\subsection{Determination of the GDC with an autocalibration approach}
Initially, we planned to solve for the GDC using an autocalibration approach in which the distortion-free reference frame is obtained by combining multiple different pointings. As the position measurements of each star are based on measurements at different parts of the detector, the effect of the GD randomises and therefore is averaged out. This is a well-proven technique and has been used to calibrate the GD of the \textit{HST} \citep{anderson_improved_2003, bellini_astrometry_2009}, but also for various ground-based studies \citep{anderson_ground-based_2006, bellini_ground-based_2010, libralato_ground-based_2014, libralato_high-precision_2015}. We refer to these publications for a detailed description of the iterative process.

\begin{figure*}
  \centering
    \includegraphics[width=0.8\textwidth]{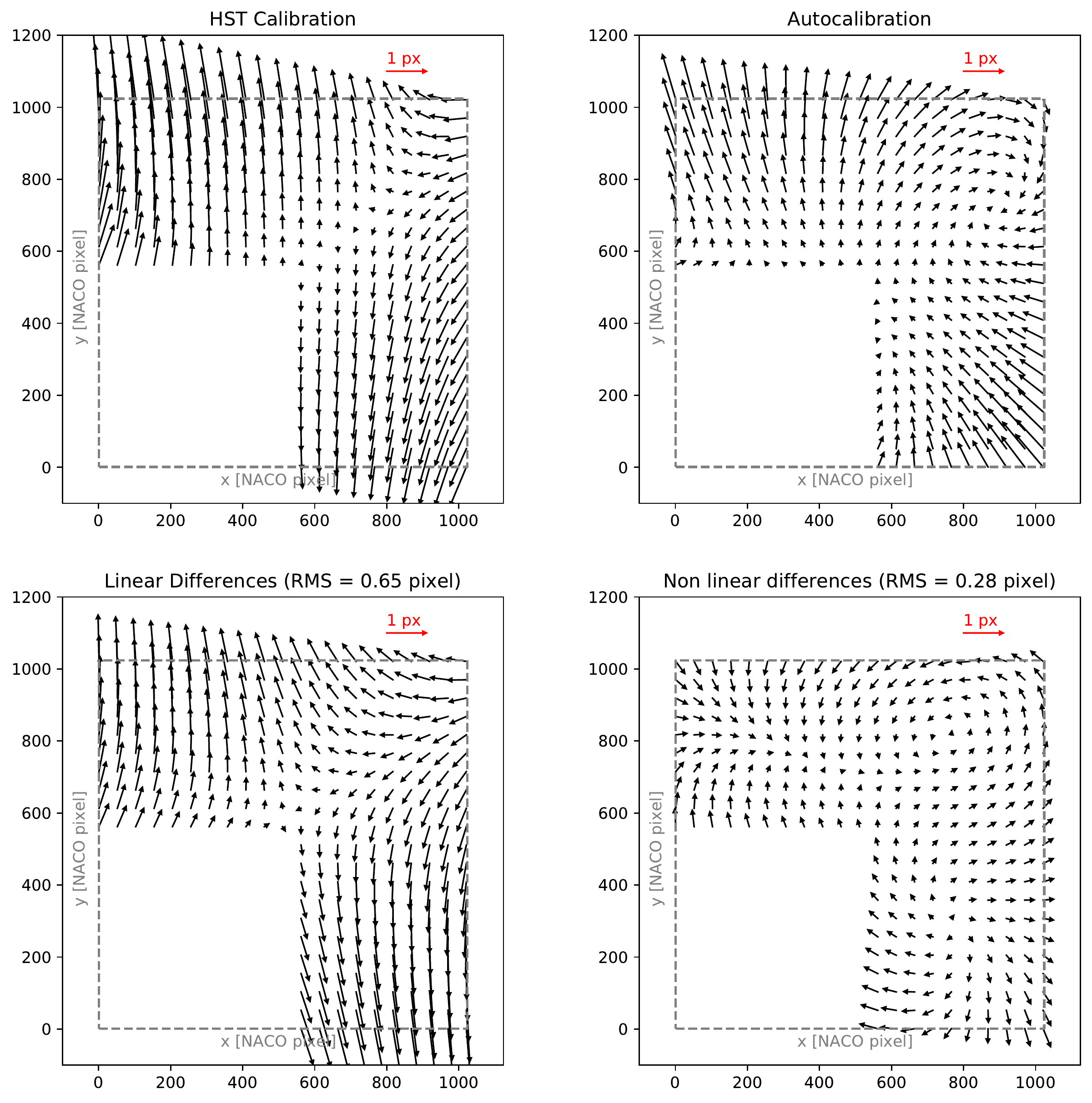}
  \caption{This figure shows a comparison of the two distortion corrections we obtained with different methods: the external calibration (upper left) versus the autocalibration (upper right). Significant differences (total differences between models: 0.68 pixels) can be observed between the maps. The deviations are dominated by a linear part (lower left, rms = 0.65~NACO~pixel), however, also the nonlinear differences are significant (lower right, rms = 0.28~NACO~pixel).}
  \label{fig:hst_vs_auto}
\end{figure*}

In comparison to the GDC obtained with an external reference (see section above) our autocalibration result showed significant differences (see \autoref{fig:hst_vs_auto}): we were unable to determine the linear distortion (the so-called skew) terms, which caused a rms difference of 0.68~NACO~pixel, but also the nonlinear differences had an rms of 0.28~NACO~pixel.

By comparing the NACO master frames with independent \textit{HST} catalogues of NGC~6441 (based on WFC3/UVIS and ACS/WFC observations), we could verify that the differences between the distortion corrections indeed are caused by uncorrected GD in the autocalibration reference frame.

It can be easily understood why we were not able to determine the linear distortion terms using the autocalibration: the linear skew terms are the same over the whole field of the detector. Even if stars are measured on different detector positions, their position measurements are affected by the same skew and therefore the reference frame also has the same skew. This degeneracy is lifted if an instrument with multiple detectors is used, as is the case for the papers cited above, or if there are pointings with different orientation on sky.

The nonlinear differences are most likely caused by a relatively low number of stars in comparison with~e.g., the autocalibration of the \textit{HST} or ground-based wide-field instruments.

The failure of the autocalibration approach with our 2018 NACO dataset highlights the importance of the choice of the dither $+$ rotation pattern. Especially if no external reference with sufficient astrometric precision is available (as will possibly be the case for future instrument at ELTs), a particular attention on this is required during the preparation of future observations.

\section{Velocity Dispersion Data}
Table \ref{data_table} contains the values of the PM-based velocity dispersions computed in our work.
\begin{table*}
\centering
\begin{tabular}{lllllll}
\hline
\begin{tabular}[l]{@{}l@{}}$R$\\ {[}arcsec{]}\end{tabular} &
  \begin{tabular}[l]{@{}l@{}}$\sigma$\\ {[}mas~yr$^{-1}${]}\end{tabular} &
  \begin{tabular}[l]{@{}l@{}}$\sigma_{\rm R}$\\ {[}mas~yr$^{-1}${]}\end{tabular} &
  \begin{tabular}[l]{@{}l@{}}$\sigma_{\rm T}$\\ {[}mas~yr$^{-1}${]}\end{tabular} & \\ \hline
  0.76 &  0.316 $\pm$  0.034 &  0.301 $\pm$  0.045 &  0.332 $\pm$  0.050\\
  1.86 &  0.305 $\pm$  0.015 &  0.315 $\pm$  0.022 &  0.294 $\pm$  0.021\\
  2.98 &  0.321 $\pm$  0.016 &  0.328 $\pm$  0.023 &  0.314 $\pm$  0.022\\
  3.91 &  0.349 $\pm$  0.018 &  0.368 $\pm$  0.025 &  0.330 $\pm$  0.023\\
  4.62 &  0.312 $\pm$  0.015 &  0.305 $\pm$  0.022 &  0.320 $\pm$  0.023\\
  5.44 &  0.278 $\pm$  0.015 &  0.261 $\pm$  0.019 &  0.293 $\pm$  0.022\\
  6.16 &  0.316 $\pm$  0.017 &  0.302 $\pm$  0.022 &  0.333 $\pm$  0.025\\
  7.01 &  0.302 $\pm$  0.016 &  0.302 $\pm$  0.022 &  0.303 $\pm$  0.022\\
  7.68 &  0.328 $\pm$  0.017 &  0.326 $\pm$  0.024 &  0.342 $\pm$  0.025\\
  8.29 &  0.287 $\pm$  0.015 &  0.262 $\pm$  0.019 &  0.295 $\pm$  0.022\\
  8.95 &  0.288 $\pm$  0.015 &  0.318 $\pm$  0.024 &  0.254 $\pm$  0.019\\
  9.98 &  0.298 $\pm$  0.016 &  0.273 $\pm$  0.020 &  0.329 $\pm$  0.024\\
 11.75 &  0.287 $\pm$  0.016 &  0.285 $\pm$  0.023 &  0.289 $\pm$  0.023\\
 \hline
\end{tabular}
\caption{Results for the proper motion dispersion profiles in the inner region of NGC~6441.}
\label{data_table}
\end{table*}


\bsp	
\label{lastpage}
\end{document}